\pdfoutput=1
\documentclass[11pt,twoside,a4paper,cmspaper,final,collab]{cms-tdr}
\def\svnVersion{506b629}\def\svnDate{2026/04/22}\def\cmsCernNoTag{CERN-EP-2025-058}\def\cmsCernDate{\today}\def\cmsMessage{Published in the Journal of High Energy Physics as \href{http://dx.doi.org/10.1007/JHEP04(2026)147}{\doi{10.1007/JHEP04(2026)147}.}}
\begin{document}\cmsNoteHeader{HIN-24-005}

\newcommand{\HeHEPData} {HEPData\xspace}
\newcommand{\rg}{\ensuremath{R_{\mathrm{g}}}\xspace}
\newcommand{\zg}{\ensuremath{z_{\mathrm{g}}}\xspace}
\newcommand{\zch}{\ensuremath{z_{\text{\PQb,ch}}}\xspace}
\newcommand{\zcut}{\ensuremath{z_{\text{cut}}}\xspace}
\newcommand{\pttrack}{\ensuremath{\pt^{\text{track}}}\xspace}
\newcommand{\ptbch}{\ensuremath{\pt^{\text{\PQb,ch}}}\xspace}
\newcommand{\ptjetch}{\ensuremath{\pt^{\text{jet,ch}}}\xspace}

\ifthenelse{\boolean{cms@external}}{\providecommand{\suppMaterial}
{the supplemental material [URL will be inserted by publisher]}}
{\providecommand{\suppMaterial}{Appendix~\ref{app:suppMat}}}

\cmsNoteHeader{HIN-24-005}
\title{Jet fragmentation function and groomed substructure of bottom quark jets in proton-proton collisions at 5.02\TeV}

\date{\today}

\abstract{
A measurement of the substructure of bottom quark jets (\PQb jets) in proton-proton ($\Pp\Pp$) collisions is presented. The measurement uses data collected in $\Pp\Pp$ collisions at $\sqrt{s} = 5.02\TeV$, with a low number of simultaneous interactions per bunch crossing, recorded by the CMS experiment in 2017, corresponding to an integrated luminosity of 301\pbinv. An algorithm to identify and cluster the charged decay daughters of \PQb hadrons is developed for this analysis, which facilitates the exposure of the gluon radiation pattern of \PQb jets using iterative Cambridge--Aachen declustering. The soft-drop-groomed jet radius, \rg, and momentum balance, \zg, of \PQb quark jets are presented. These observables can be used to test perturbative quantum chromodynamics predictions that account for mass effects. Because the \PQb hadron is partially reconstructed from its charged decay daughters, only charged particles are used for the jet substructure studies. In addition, a jet fragmentation function, \zch, is measured, which is defined as the distribution of the ratio of the transverse momentum (\pt) of the partially reconstructed \PQb hadron with respect to the charged-particle component of the jet \pt. The substructure variable distributions are unfolded to the charged-particle level. The \PQb jet substructure is compared to the substructure of jets in an inclusive jet sample that is dominated by light-quark and gluon jets in order to assess the role of the \PQb quark mass. A strong suppression of emissions at small \rg values is observed for \PQb jets when compared to inclusive jets, consistent with the dead-cone effect. The measurement is also compared with theoretical predictions from Monte Carlo event generators. This is the first substructure measurement of \PQb jets that clusters together the \PQb hadron decay daughters independent of the \PQb hadron species and decay channel.}

\hypersetup{%
pdfauthor={CMS Collaboration},%
pdftitle={Jet fragmentation function and groomed substructure of bottom quark jets in proton-proton collisions at 5.02 TeV},%
pdfsubject={CMS},%
pdfkeywords={CMS, measurement, jet substructure, quarks, gluons, jets, Lund plane}} 

\maketitle 

\section{Introduction}\label{sec:introduction}

Jets are the collimated showers of particles that result from the fragmentation of energetic quarks and gluons produced in high-energy particle collisions.
They are one of the main objects used to improve our understanding of the strong nuclear force, which is described by quantum chromodynamics (QCD). 
Describing the formation of jets requires a detailed understanding of parton showering, \ie, the cascade of partons initiated by a highly energetic parton produced in the hard scattering, and the strongly coupled transition from partons into hadrons, known as hadronization. The radiation pattern of jets contains valuable information about their formation, which can be analyzed through jet substructure observables~\cite{Larkoski:2017jix, Marzani:2019hun, Kogler:2021kkw}.

At a first approximation, the jet shower initiated by light-flavor quarks (up, down, or strange quarks) and gluons is governed by the infrared and collinear divergences of QCD and the running of the strong coupling, \alpS. For heavy-flavor quark jets, \ie, those initiated by charm (\PQc jets) or bottom quarks (\PQb jets), there are no such divergences due to the heavy-quark masses, which effectively regularize them. In the quasi-collinear approximation, the Dokshitzer--Gribov--\linebreak
Lipatov--Altarelli--Parisi (DGLAP) splitting function for a gluon emission off a heavy quark (${\PQQ\to \PQQ\Pg}$) is given by
\begin{equation}
    \label{eq:massive_splitting}
    P_{\PQQ\to \PQQ\Pg}(z) = \frac{1-z}{z} + \frac{z}{2} - 2 \mu^2_{\PQQ\Pg},
\end{equation}
where $z$ is the momentum fraction of the emission, and the mass ratio $\mu^2_{\PQQ\Pg}$ is given by
\begin{equation}
    \mu^2_{\PQQ\Pg} = \frac{m_{\PQQ}^2}{m_{\PQQ\Pg}^2 - m_{\PQQ}^2},
\end{equation}
where $m_{\PQQ}$ is the mass of the heavy quark and $m_{\PQQ\Pg}$ is the invariant mass of the heavy-quark and gluon system~\cite{Gribov:1972ri, Dokshitzer:1977sg, Altarelli:1977zs, Catani:2000ef, Ilten:2017rbd}. 
Equation (\ref{eq:massive_splitting}) has no soft divergence and it reduces to the standard DGLAP splitting function in the massless limit $m_{\PQQ} \to 0$. 
As a consequence of the mass term, the momentum balance of a gluon emission off a heavy quark is expected to be more asymmetric in energy than it is for light quarks ($\PQq\to \PQq\Pg$), or gluons ($\Pg\to \Pg\Pg$), such that the heavy quark carries a larger amount of the momentum of the branching. This results in the heavy-flavor hadron that is observed experimentally carrying a substantial fraction of the heavy-quark momentum. 
The mass term also leads to the ``dead-cone'' effect, a general feature of quantum gauge field theories whereby collinear radiation is suppressed for massive particles. 
In QCD, the presence of a mass term in the quark propagator leads to a suppression of gluon emissions at a small angle relative to the recoiling heavy quark~\cite{Dokshitzer:1991fd}.
The probability to have a gluon emission with an opening angle $\theta$ with respect to the heavy quark in a $\PQQ \to \PQQ\Pg$ branching is given by
\begin{equation}
\label{eq:deadCone}
    \rd \mathcal{P}(\theta) \propto \frac{\rd \theta^2 }{(\theta^2+\theta_0^2)^2},
\end{equation}
where the parameter ${\theta_0 = m_{\PQQ}/E_{\PQQ}}$ dictates the effective size of the dead-cone angle, with $E_{\PQQ}$ being the energy of the heavy quark before the branching. 
The resulting dead-cone region is larger for more massive quarks or for a gluon emission from a less energetic heavy quark. 
In the massless limit, ${\theta_{0} \to 0}$, one recovers the textbook collinear divergence of massless QCD. 
The dead-cone effect has been directly observed by the ALICE Collaboration using jets containing a fully reconstructed \PDz meson~\cite{ALICE:2021aqk}.
More recently, the LHCb Collaboration has made a measurement of the Lund jet plane of jets containing a \PBpm meson, accessing the \PQb quark dead cone via a comparison to a light-quark-enriched jet sample of jets recoiling off of \PZ bosons~\cite{LHCb:2025mcq}.

Heavy-quark fragmentation is a rich, multiscale process. It is initiated by the highly energetic parton, carrying a transverse momentum (\pt) of the order of hundreds of \GeV. It then evolves with the relative transverse momentum (\kt) of the $1\to 2$ branchings in the parton shower,  which ranges from a few \GeV up to the order of the jet \pt. It also involves the mass of the heavy quark, which is about 1.5 and 4\GeV for charm and bottom quarks, respectively, and concludes with the energy scale associated with the transition from partons to hadrons, ${\Lambda_\text{QCD} \approx 200 \MeV}$. The description of all of these effects in Monte Carlo (MC) event generators is not trivial. Thus, precision measurements that can constrain such aspects of heavy-flavor jet fragmentation can help shed light on them. In this analysis, the focus is on \PQb jets, where mass effects are the strongest and for which CMS has developed tools for their identification (``tagging'')~\cite{CMS:2017wtu}. Understanding the radiation pattern of \PQb jets is also important to improve the description of top quark decays, as well as Higgs decays to bottom quark-antiquark pairs~\cite{Larkoski:2017jix, Kogler:2018hem, Marzani:2019hun}. 
In addition, \PQb jets are of interest for the characterization of the quark-gluon plasma properties in heavy ion collisions~\cite{CMS:2013qak, 
 ATLAS:2022agz, CMS:2018dqf, CMS:2024krd}. 
 In this context, heavy-flavor jets introduce an additional energy scale in the multiscale process and can be used to effectively constrain the color factor of the jet shower. 
 The dead-cone region has been identified as a phase space region where medium-induced radiation effects can be studied in a controlled environment~\cite{Cunqueiro:2018jbh, Cunqueiro:2022svx}.
 Previous measurements of the differential jet shape in proton-proton ($\Pp\Pp$) and Pb-Pb collisions~\cite{CMS:2022btc} have indicated differences between the light-flavor jet population and \PQb quark jet population, as well as their modifications due to medium effects, with hints of the dead-cone effect being at play.

Interpreting the data in terms of the underlying parton branching process becomes more challenging when \PQb hadron decay daughters are treated on an equal footing with hadrons from \PQb quark hadronization. This challenge arises, for example, in Lund-plane-based substructure observables or energy-energy correlators~\cite{Cunqueiro:2018jbh, Lee:2019lge}, but it generally affects jet substructure observables.
A recent reinterpretation of DELPHI and OPAL electron-positron ($\Pep\Pem$) data of light-hadron fragmentation functions measured in heavy-quark-enriched events~\cite{DELPHI:2011aa,OPAL:2002plk} showed that, after performing a simulation-based subtraction of the contribution of charm or bottom hadron decays, the resulting light-hadron fragmentation function is directly sensitive to the dead-cone effect \cite{Kluth:2023umf}. 
Thus, a strategy to identify and isolate the decay daughters of heavy-flavor hadrons must be adopted in order to have a reliable interpretation of the data in terms of the underlying parton branching process. 
This approach is employed in this analysis and is discussed in more detail in Section~\ref{sec:method}.

In this analysis, we present a measurement of three jet substructure observables.
Each of them probes different aspects of the fragmentation pattern of jets. 
Their technical definition is discussed in Section~\ref{sec:observables}. 
The first two observables are the groomed jet radius, \rg, and momentum balance, \zg.
They are obtained using the soft-drop (SD) grooming algorithm \cite{Larkoski:2014wba}, which is a generalization of the modified mass-drop algorithm~\cite{Dasgupta:2013ihk}. 
The SD algorithm allows for the decomposition of a full jet into a hard two-prong subjet structure, using iterative Cambridge--Aachen (CA) declustering~\cite{Dokshitzer:1997in}. 
The momentum balance \zg is the ratio of the transverse momentum carried by the softer subjet over the sum of the momenta of the two subjets, which is, to first approximation, related to the DGLAP splitting function. 
The groomed jet radius \rg is the angular opening between the softer and harder subjets.
It is used as a proxy for the separation of the gluon emissions with respect to the \PQb quark. 
Because of the dead-cone effect, small \rg emissions are expected to be suppressed. Recent theoretical calculations achieve next-to-leading logarithmic accuracy for the \zg and \rg distributions of heavy-flavor jets \cite{Caletti:2023spr}. 
The third observable considered in the measurement is a jet fragmentation function, namely,  the distribution of the fraction of the charged-particle transverse momentum of the jet carried by the charged-particle decay products of the \PQb hadron, ${\zch = \ptbch/\ptjetch}$. 
The fragmentation function for \PQb hadrons was constrained at LEP~\cite{DELPHI:2011aa, OPAL:2002plk, ALEPH:2001pfo}, but recent CERN LHC measurements \cite{ATLAS:2021agf, ATLAS:2022miz} show that there are deviations with respect to predictions provided by MC event generators. 
These deviations suggest that there is a possible break of universality, likely resulting from the different color field environments of $\Pp\Pp$ and $\Pe^+\Pe^-$ collisions, which the present measurement can help clarify further. 
At leading order (LO) in perturbation theory, \zch and \zg should be strongly correlated, but such correlations are lost when accounting for higher-order and nonperturbative corrections~\cite{Caletti:2023spr}. Thus, \zch offers a complementary handle to the \zg observable.
The jet substructure observables are measured both for inclusive jets and \PQb jets and used to constrain the fragmentation of quark and gluon showers, providing important constraints to MC event generators and theoretical calculations.

The measurement is performed using data from $\Pp\Pp$ collisions at ${\sqrt{s} = 5.02\TeV}$ recorded by the CMS experiment in 2017. 
The CMS detector components, along with the reconstruction algorithms used, are described in Section~\ref{sec:cms_detector}.
Details about the measured and simulated samples are presented in Section~\ref{sec:data_samples}. The observable definitions and the strategy for the corrections are described in Sections~\ref{sec:observables} and~\ref{sec:method}, respectively. 
The systematic uncertainties associated with the corrected distributions are described in Section~\ref{sec:syst_uncertainties}. 
The results are discussed in Section~\ref{sec:results}. 
A summary of the measurement is presented in Section~\ref{sec:summary}.
Tabulated results are provided in the HEPData record for this analysis~\cite{hepdata}.

\section{The CMS detector and event reconstruction}\label{sec:cms_detector}

The central feature of the CMS apparatus is a superconducting solenoid of 6\unit{m} internal diameter, providing a magnetic field of 3.8\unit{T}. Within the magnetic volume are a silicon pixel and strip tracker, a lead tungstate crystal electromagnetic calorimeter (ECAL), and a brass and scintillator hadron calorimeter (HCAL), each composed of a barrel and two endcap sections. Forward calorimeters extend the pseudorapidity ($\eta$) coverage provided by the barrel and endcap detectors. Muons are measured in gas-ionization detectors embedded in the steel flux-return yoke outside the solenoid. A more detailed description of the CMS detector, together with a definition of the coordinate system used and the relevant kinematic variables, is presented in Refs.~\cite{CMS:2008xjf, CMS:2023gfb}.

The primary vertex (PV) is taken to be the vertex corresponding to the hardest scattering in the event, evaluated using tracking information alone, as described in Section 9.4.1 of Ref.~\cite{CMS-TDR-15-02}.

The particle-flow algorithm~\cite{CMS-PRF-14-001} aims to reconstruct and identify each individual particle in an event, with an optimized combination of information from the various elements of the CMS detector. The energy of photons is obtained from the ECAL measurement. The energy of electrons is determined from a combination of the electron momentum at the primary interaction vertex as determined by the tracker, the energy of the corresponding ECAL cluster, and the energy sum of all bremsstrahlung photons spatially compatible with originating from the electron track. The energy of muons is obtained from the curvature of the corresponding track. The energy of charged hadrons is determined from a combination of their momentum measured in the tracker and the matching ECAL and HCAL energy deposits, corrected for the response function of the calorimeters to hadronic showers. Finally, the energy of neutral hadrons, not associated with tracks, is obtained from the corrected ECAL and HCAL energies.

For each event, jets are clustered from all reconstructed particles in the event using the anti-\kt algorithm~\cite{Cacciari:2008gp} with ${R = 0.4}$, as implemented in \textsc{FastJet}~\cite{Cacciari:2011ma}.
Jet momentum is determined as the vectorial sum of all particle momenta in the jet, and based on simulation, is found to be, on average, within 5 to 10\% of the true momentum over the whole \pt spectrum and detector acceptance. Additional $\Pp\Pp$ interactions within the same or nearby bunch crossings (pileup) can contribute additional tracks and calorimetric energy depositions, increasing the apparent jet momentum. To mitigate this effect, tracks identified to be originating from pileup vertices are discarded and an offset correction is applied to correct for remaining contributions~\cite{CMS:2020ebo}. Jet energy corrections are derived from simulation studies so that the average measured energy of jets becomes identical to that of particle-level jets. In situ measurements of the momentum balance in dijet, photon$+$jet, $\PZ+$jet, and multijet events are used to determine any residual differences between the jet energy scale in data and in simulation, and appropriate corrections are made~\cite{CMS:2016lmd}. 
The jet energy corrections used for this analysis were derived from a high-luminosity pp data sample at ${\sqrt{s}=13\TeV}$ collected also in 2017. 
We verify in the simulated samples used for this analysis that, in the kinematic range under study, the corrected jet \pt agrees within 0.2\% with the generated one.
Additional selection criteria are applied to each jet to remove jets potentially dominated by instrumental effects or reconstruction failures~\cite{CMS:2020ebo}.
The jet energy resolution amounts typically to 10\% at 100\GeV~\cite{CMS:2016lmd}.

Jets containing one or more \PQb hadrons are identified using \textsc{ParticleNet}, a jet tagging tool based on the graph neural network architecture~\cite{Qu:2019gqs}.
The \PQb-tagging model, trained on high-pileup $\Pp\Pp$ collisions at 13\TeV, performs comparably on the low-energy, low-pileup simulated sample and requires no retraining.
A selection on the \textsc{ParticleNet} score is applied to select \PQb jets with a 52\% efficiency and 0.01\% misidentification rate according to the low-energy and low-pileup simulation.

Track reconstruction relies on the silicon tracker, which was upgraded at the start of 2017 with the installation of a new pixel detector~\cite{Phase1Pixel}. The upgraded tracker measures particles up to $\abs{\eta} = 3.0$ with typical resolutions of 1.5\% in \pt and 20--75\mum in the transverse impact parameter~\cite{DP-2020-049} for nonisolated particles of ${1<\pt<10\GeV}$. 
Tracks used in this analysis are required to have ${\abs{\eta}<2.4}$ and ${\pt>1\GeV}$. 
The $\eta$ selection is chosen such that the \PQb tagging performance is consistently high, aiming to reduce data-to-simulation differences in the clustering of the \PQb hadron decay daughters. 
The ${\pt>1\GeV}$ requirement is to avoid the fast drop of the track reconstruction efficiency at low \pt. 
To further discard tracks originating from pileup vertices, the absolute value of the longitudinal and transverse impact parameter of each track is required to be smaller than 17 and 0.2 \cm, respectively~\cite{CMS:2017wtu}. 

Events of interest are selected using a two-tiered trigger system. The first level, composed of custom hardware processors, uses information from the calorimeters and muon detectors to select events at a rate of around 100\unit{kHz} within a fixed latency of about 4\mus~\cite{CMS:2020cmk}. 
The second level, known as the high-level trigger, consists of a farm of processors running a version of the full event reconstruction software optimized for fast processing, and reduces the event rate to around 1\unit{kHz} before data storage~\cite{CMS:2016ngn}. 

\section{Data and simulated samples}\label{sec:data_samples}

This analysis uses $\Pp\Pp$ collision data collected by the CMS experiment in 2017 at ${\sqrt{s} = 5.02\TeV}$, corresponding to an integrated luminosity of $301\pm 6\pbinv$ \cite{CMS-PAS-LUM-19-001}. The average pileup activity during this run is approximately two interactions per bunch crossing. Events with high-\pt jets are collected with triggers requiring at least one jet with a \pt above 40, 60, 80, or 100\GeV. Because of the low pileup activity during the 5.02\TeV $\Pp\Pp$ run, the 100\GeV trigger is able to sample the full luminosity. However, the lower threshold triggers only collected a part of the total luminosity, which affects the trigger efficiency. The latter is found to be about 99\% for jets with an offline reconstructed \pt of 80\GeV, which is the lowest jet \pt value considered in this measurement.
There are about two million jets satisfying the \PQb tagging requirements described in Section \ref{sec:method} in the kinematic range considered.

Simulated events are used for the extraction of corrections and the estimation of systematic uncertainties on the measurement. The nominal simulated sample consists of QCD multijet events, generated at LO using \PYTHIA{8.240}~\cite{Sjostrand:2014zea}, with the hadronization and underlying event parameters set to the CP5 tune~\cite{CMS:2019csb}. The \PYTHIA{8} event generator implements a dipole parton shower ordered in \pt. The gluon radiation off of heavy quarks is taken into account via matrix element corrections~\cite{Norrbin:2000uu,Bierlich:2022pfr}. The quark and gluon hadronization is described by the Lund string model~\cite{Andersson:1983ia,Sjostrand:1984iu}, with the Bowler~\cite{Bowler:1981sb} modification of the fragmentation function for heavy quarks. 

A second simulation sample is produced, also at LO, using \HERWIG{7.1.4}~\cite{Bellm:2017bvx} with the CH3 tune~\cite{Sirunyan:2020pqv}. In this case, the parton shower is governed by angular-ordered radiation~\cite{Gieseke:2003rz}, with mass-dependent splitting functions~\cite{Marchesini:1989yk,Gieseke:2003rz}. The transition to hadrons is described by the cluster fragmentation model~\cite{Webber:1983if}.

The generated samples use the next-to-next-to-LO NNPDF 3.1~\cite{Ball:2014uwa} parton distribution functions with $\alpS(m_\PZ) = 0.118$, where $m_{\PZ}$ represents the \PZ boson mass. Both generated samples are processed through a detailed simulation of the CMS detector with the \GEANTfour toolkit~\cite{Agostinelli:2002hh}. Heavy-quark mass effects are modelled in the parton showers of \PYTHIA{8} CP5 and \HERWIG{7} CH3, allowing for an accurate description of soft gluon emission from heavy quarks over a wide angular region, including the collinear direction.

\section{Observable definition}\label{sec:observables}

\subsection{Groomed observables}
Two of the substructure observables considered in this measurement are derived from the SD grooming algorithm \cite{Larkoski:2014wba}.
This algorithm reclusters the original anti-\kt jet constituents using the CA algorithm \cite{Dokshitzer:1997in}, which is a pairwise clustering of particles (and subjets thereafter) that clusters the pair of subjets with the smallest rapidity-azimuth distance at each step of the clustering process. No explicit selection on the $\Delta R$ of the constituents with respect to the jet axis is applied for the reclustering step.
This strong angular ordering of the CA algorithm intends to mimic the approximate angular ordering expected from color coherence effects in QCD showers~\cite{Dokshitzer:1997in, Wobisch:1998wt}. Once the CA clustering tree is constructed, the clustering history is followed in reverse, always following the hardest branch at each step of the declustering process, until the SD condition is satisfied:
\begin{equation}
    \label{eq:softDrop}
    \frac{\text{min}(p_{\mathrm{T},1}\text{,} p_{\mathrm{T},2})}{p_{\mathrm{T},1} + p_{\mathrm{T},2}} > \zcut \left( \frac{\Delta R_{1,2}}{R}\right)^\beta,
\end{equation}
where $p_{\mathrm{T,i}}$ are the \pt of each subjet and $\Delta R_{1,2}$ is the distance on the rapidity-azimuth plane of the two subjets at each declustering step, \ie, $\Delta R_{1,2}^2 = \Delta y_{1,2}^2 + \Delta \phi_{1,2}^2$.
The parameter $R$ is the jet clustering distance parameter ($R=0.4$ in this case), and \zcut and $\beta$ are the SD grooming parameters, which are tunable. The softer subjet at each step of the declustering process may be considered as a proxy for a parton emission in the jet shower.
When Eq.~(\ref{eq:softDrop}) is satisfied, the CA declustering iteration is stopped. The two resulting subjets are then used as proxies for the hard two-prong structure of the jet. 
The observables of interest are the groomed jet radius ${\rg = \Delta R_{1,2}}$ and the groomed momentum balance ${\zg = p_{\mathrm{T},2}/(p_{\mathrm{T},1} + p_{\mathrm{T},2})}$, where $p_{\mathrm{T},2}$ is the transverse momentum of the softer subjet.
The  groomed jet radius \rg is presented in the logarithmic form $\ln(R/\rg)$ in all the figures presented in this paper, following the standard representation for observables based on the CA clustering tree.

The grooming parameters \zcut and $\beta$ determine the degree to which soft- and wide-angle radiation is being removed by the algorithm. Increasing the value of \zcut preferentially removes soft radiation, whereas decreasing values of $\beta$ preferentially remove wide-angle radiation. We use the SD grooming parameters of $\beta = 0$ and $\zcut = 0.1$. The value of $\zcut = 0.1$ is the standard value used at the LHC, which represents a sufficiently low value to remove soft and wide-angle radiation. The choice of $\beta = 0$, together with $\zcut = 0.1$, facilitates the comparison with theoretical calculations of heavy-flavor quark substructure ~\cite{Caletti:2023spr}. In addition, the suppression of emissions due to the dead-cone effect is clearest for $\beta = 0$, as illustrated in Fig.~3 of Ref.~\cite{Caletti:2023spr}.

Jets that fail the SD condition are referred to as ``one-pronged'' or SD-untagged jets. The population of such jets is tracked in the analysis with an additional bin used for the unfolding corrections, as described later in the paper.

\begin{figure}[ht]
    \centering
    \includegraphics[width=0.99\linewidth]{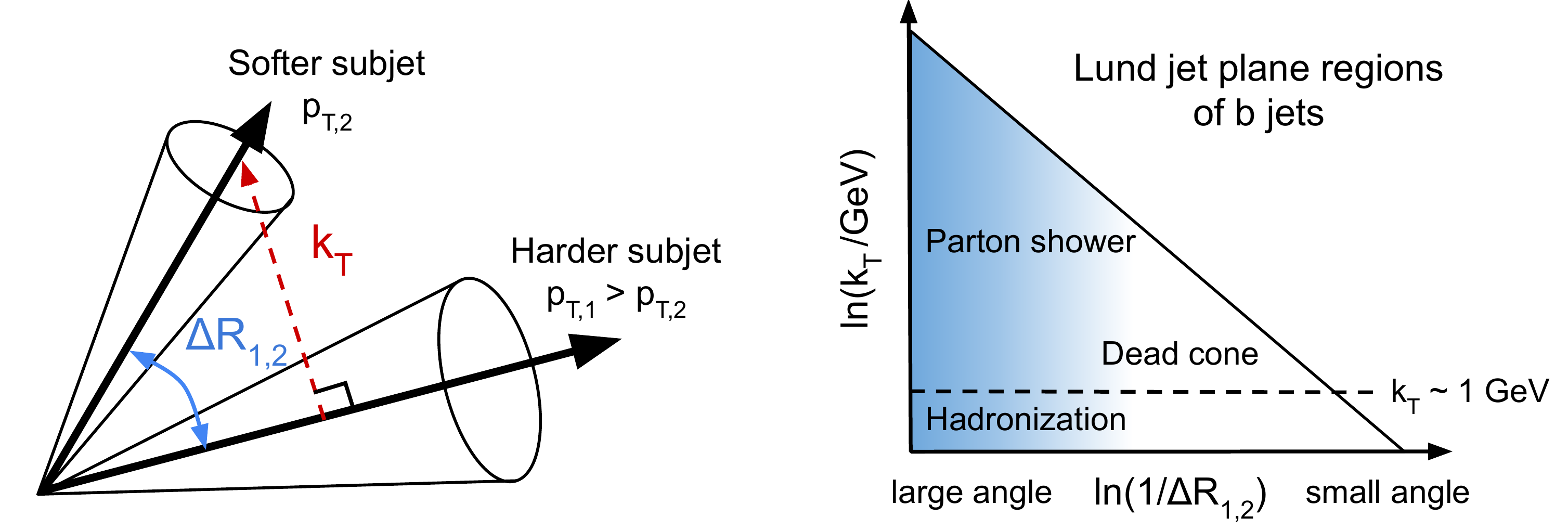}
    \caption{Left panel: A schematic diagram of two Cambridge--Aachen subjets, such as those found with the soft-drop grooming algorithm. The splitting angle $\Delta R_{1,2}$ and the relative transverse momentum \kt between the two subjets are annotated. Right panel: The Lund jet plane regions of bottom quark jets. The vertical axis is the logarithm of the relative transverse momentum $\kt/\GeV$ of the softer subjet with respect to the harder subjet. The horizontal axis is the logarithm of the inverse of the opening angle between the softer and harder subjets, $\Delta R_{1,2}$. The Lund jet plane is expected to be dominated by hadronization effects for \kt below the \GeV scale. Above the \GeV scale, the Lund jet plane provides information on the parton showering description. Emissions are suppressed at small angles due to the dead-cone effect. The \PQb hadron decays (not depicted) populate the same region. The blue shading, fading from left to right, represents the density of emissions in the primary Lund jet plane for \PQb quark jets.}
    \label{fig:bjet_lp}
\end{figure}

{\tolerance=800
The relative transverse momentum of the softer subjet with respect to the harder subjet ${\kt = p_{\mathrm{T},2}\rg}$, which is a measure of the energy scale of the branching, is used as an additional selection requirement on the SD emissions considered in the analysis. We select SD-groomed emissions that satisfy $\kt > 1\GeV$, which is a minimal momentum scale where hadronization effects are less strong~\cite{Dreyer:2018nbf, Cunqueiro:2018jbh}. As pointed out in Ref.~\cite{Cunqueiro:2018jbh}, such a selection requirement improves the hadron-to-parton correspondence of the substructure of jets and enhances the visibility of the dead-cone effect at small \rg values. 
The jets that satisfy the SD condition, but where the emission has $\kt<1\GeV$, are grouped together with the population of jets that do not satisfy the SD grooming criteria.
\par}

In \PYTHIA{8} CP5, around 3\% of inclusive jets fail the SD grooming criteria, while approximately 37\% of inclusive jets satisfy the SD condition but with $\kt<1\GeV$. For \PQb jets, these fractions are substantially different,  46\% and 25\%, respectively, due to the suppressed collinear radiation and hard fragmentation of the \PQb quark.

Groomed jet observables \cite{ALICE:2021mqf, ATLAS:2019mgf, CMS:2021iwu, CMS:2018vzn, ATLAS:2019kwg, ATLAS:2024wrd}, as well as other observables based on the CA clustering tree~\cite{ATLAS:2020bbn, CMS:2023lpp, ATLAS:2024dua}, have been measured before for inclusive jet collections.
In the case of charm quark jets, jets containing a \PDz meson have been used for substructure measurements by the ALICE and CMS Collaborations \cite{ALICE:2022phr,ALICE:2021aqk,CMS:2025afq}.
However, the approach based on exclusive decays is much more limiting in the case of \PQb hadrons, since the fraction of b hadrons that decay exclusively into charged particles is of the order of 0.1\% \cite{PDG2024}.
As a result, different LHC measurements of \PQb jet substructure so far \cite{CMS:2018ypj, CMS:2020geg} have not taken into account the effect of the decay of \PQb hadrons, \ie, the decay daughters are treated equally with the hadrons that are produced from the transition from partons to hadrons in the hadronization process. It is therefore challenging to disentangle mass effects, such as the presence of the dead cone, or to separate the parton cascade from the strong interactions from the \PQb hadron decays, which are primarily associated with weak-interaction decays.

\begin{figure}[ht]
    \centering    \includegraphics[width=\textwidth]{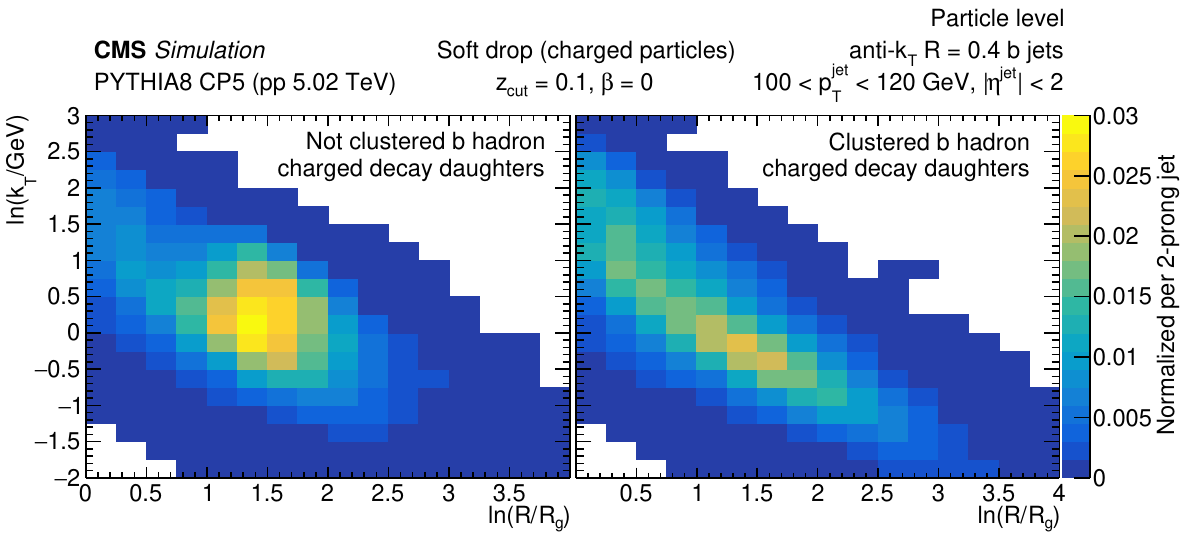}
    \caption{The Lund jet plane of the soft-drop emissions of \PQb quark jets at the particle-level of the \PYTHIA{8} CP5 simulation. The vertical axis is the logarithm of the $\kt/\GeV$ of the emission, whereas the horizontal axis is the logarithm of $R$/\rg, such that large-angle emissions populate the left-hand side of the diagrams and small-angle emissions populate the right-hand side. On the left-hand side, the SD algorithm is applied to the charged-particle jet constituents including the \PQb hadron decay daughters. On the right-hand side, the \PQb hadron charged decay daughters have been clustered together prior to the SD grooming, \ie, the charged component of the \PQb hadron remains intact. The effect of the decays can be observed in the yellow hotspot at small angles and low \kt on the left.}
    \label{fig:rgkt_gen}
\end{figure}

To visualize the effect of the \PQb hadron decays, it is instructive to use a representation of the substructure of the jet that accounts for momentum and angular information where different jet showering effects can be separated. This can be done, for example, via the Lund jet plane \cite{Dreyer:2018nbf}, which is a two-dimensional representation of the phase space of intrajet $1\to 2$ emissions constructed using the CA clustering tree. The vertical axis of this plane corresponds to the logarithm of $\kt/\GeV$ and the horizontal axis is the logarithm of the inverse of the opening angle $\Delta R_{1,2}$. The Lund jet plane allows for the identification of regions that are dominated by the short-distance physics (parton shower) and long-distance physics (hadronization) depending on the \kt region of interest. The Lund jet plane enables the separation of different effects in a modular fashion, such that at low \kt (approximately below 1\GeVns) it is dominated by hadronization effects, at large angles it is dominated by the parton showering description, and at small angles one can isolate the suppression of emissions expected from the dead-cone effect. In Fig.~\ref{fig:bjet_lp} (left) we give a schematic representation of two subjets, annotating the splitting angle and relative transverse momentum between them, while in Fig.~\ref{fig:bjet_lp} (right) we observe the different regions of the \PQb jet showers at the hadron level, assuming that there are no \PQb hadron decays. 

By analyzing the kinematics of the SD emissions in simulation in terms of the Lund jet plane, as in Fig.~\ref{fig:rgkt_gen}, we observe that the effect of the \PQb hadron decays on the substructure observable distributions is significant. 
The Lund jet plane of SD emissions of \PQb jets is highly influenced by the presence of decay products (left-hand-side distribution of Fig.~\ref{fig:rgkt_gen}) or by their absence (right-hand-side distribution of Fig.~\ref{fig:rgkt_gen}).
Left untreated, the \PQb hadron decays degrade the correspondence between the jet substructure variables and the underlying QCD splitting processes.
It is therefore necessary to address the heavy-flavor hadron decays in order to make any meaningful conclusions about the hard substructure of heavy-flavor jets. 

\subsection{Partial  \texorpdfstring{\PQb}{b} hadron reconstruction}

In this analysis, to mitigate the distortions caused by the \PQb quark decay daughters, an algorithm that reconstructs the \PQb hadron from its charged decay products is developed. This algorithm treats inclusively all possible decay channels of the \PQb hadron.
A multivariate analysis is performed to identify charged particles originating from the \PQb hadron decay using events generated by \PYTHIA{8} CP5.
Once all charged particles in the jet have been classified, the assumed decay product four-momenta are summed vectorially and such decay product particles are replaced by the partially reconstructed \PQb hadron four-momentum. 
The CA reclustering and declustering process is then applied on the new set of jet constituents, which no longer includes the decay products.

\begin{figure}[ht]
    \centering
    \includegraphics[width=0.49\textwidth]{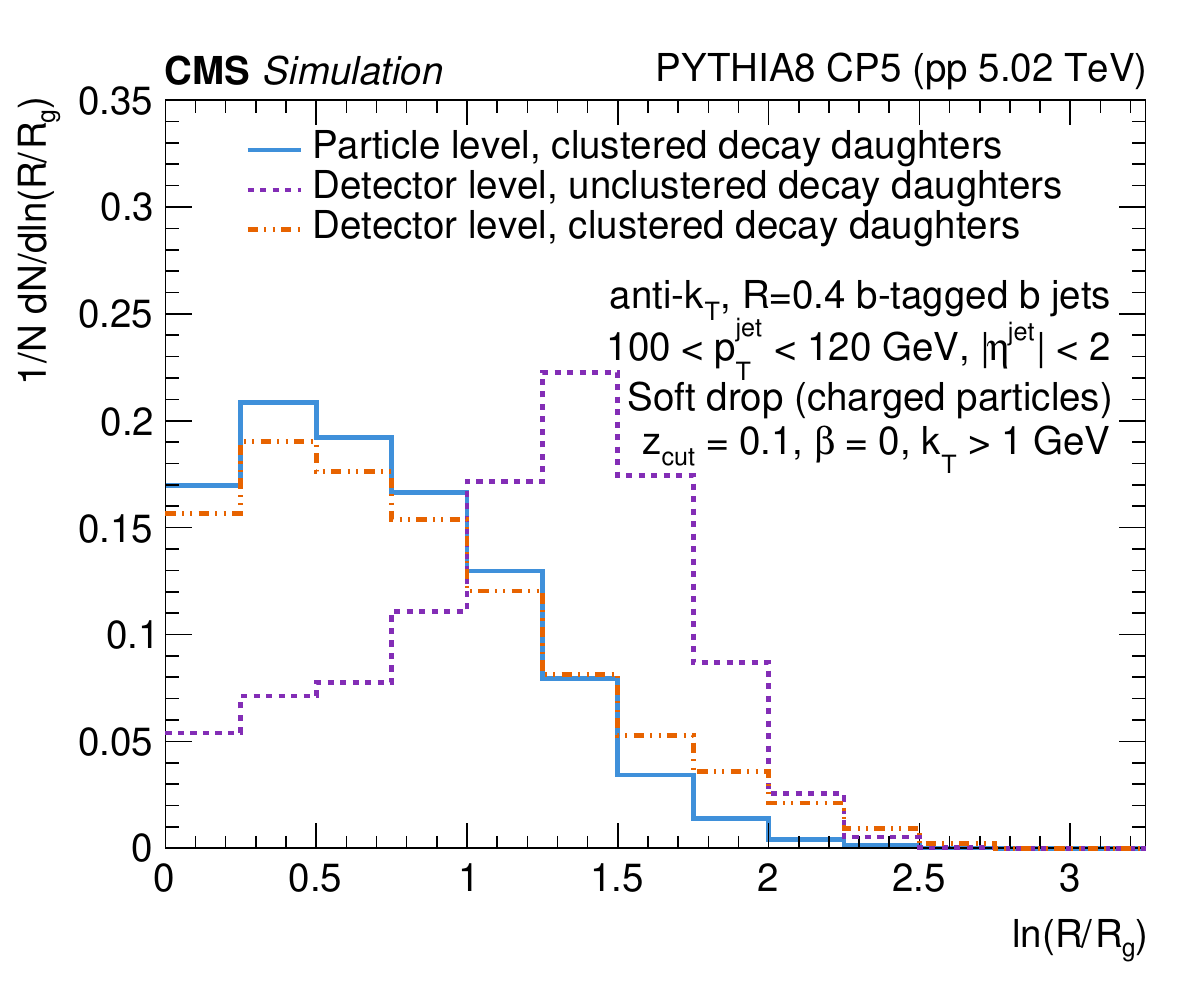}
    \includegraphics[width=0.49\textwidth]{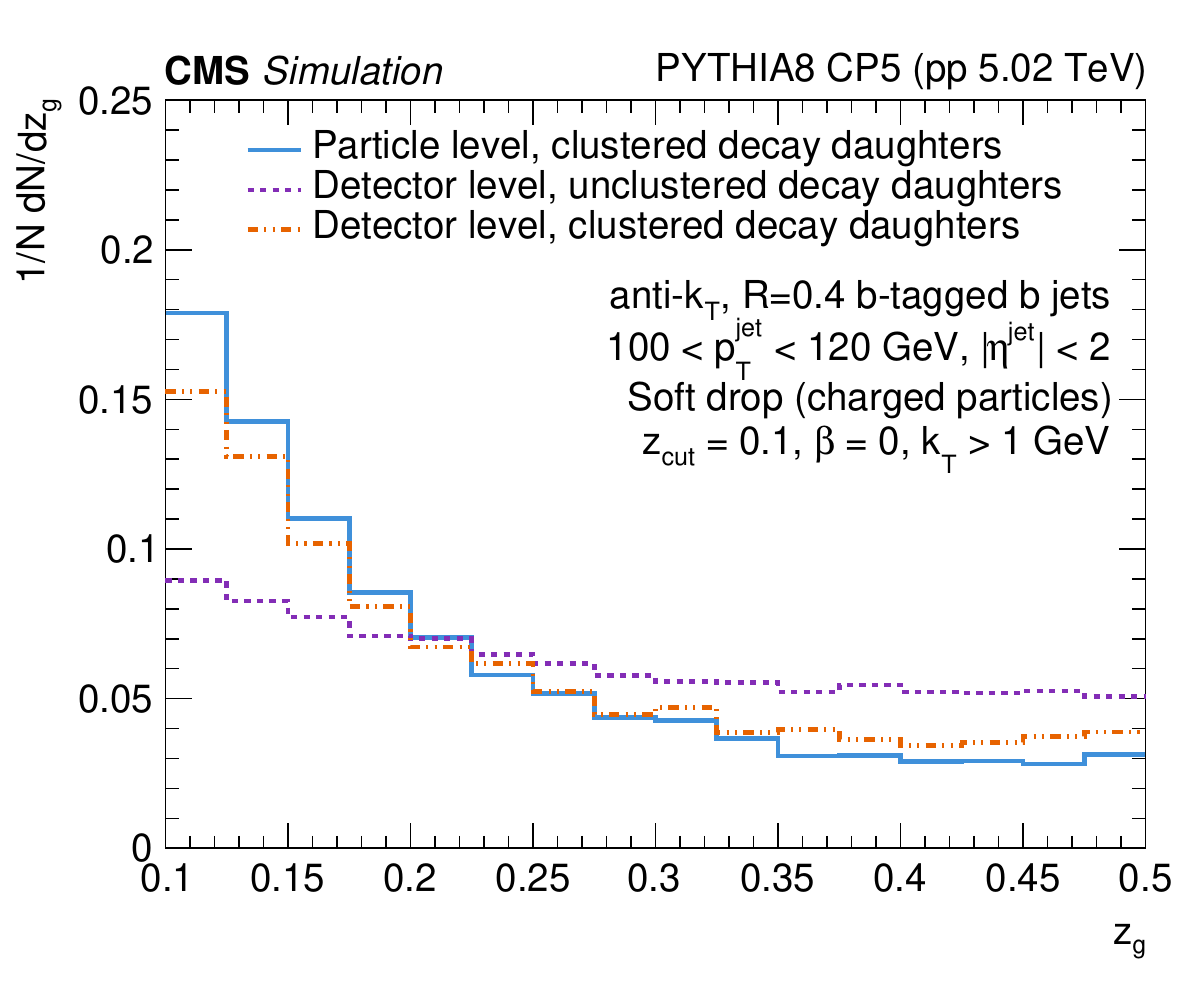}
    \caption{The modification of the \rg (left) and \zg (right) distributions at the detector level without (violet dotted curve) and with (orange dashed-dotted curve) the partial reconstruction of the \PQb hadron, compared to the particle-level distribution with the charged part of the generated \PQb hadron intact (blue solid curve). The events are produced by the \PYTHIA{8} CP5 generator.}
    \label{fig:bdt_roc}
\end{figure}

The model used in the analysis is a gradient boosted decision tree.
The input variables are characteristics of the charged particle, such as its reconstructed track and associated secondary vertex information.
The charged particles used in the training and validation of the model have ${\pttrack>1\GeV}$ and come from \PQb-tagged jets with ${\pt^{\text{jet}}>30\GeV}$. All reconstructed particles considered are matched to generated particles, so that they either come from the primary vertex (background) or the \PQb hadron decay (signal). The working point selected corresponds to 92\% signal efficiency and 95\% background rejection rate according to simulation. The performance of the classifier can also be showcased in the simulated detector-level distributions of the groomed observables \rg and \zg, as presented in the left and right panels of Fig.~\ref{fig:bdt_roc}, respectively. The distributions including the partially reconstructed \PQb hadron (orange dashed-dotted curve) approach the particle-level distribution (blue solid curve) significantly more than in the case where the \PQb hadron decay daughters are not clustered (violet dotted curve).

With the algorithm described in this section, the \PQb hadron can be partially reconstructed using charged particles. Thus, using neutral particles in addition to the charged ones for the jet substructure observables can lead to \PQb hadron neutral decay daughters populating the small angle region of the Lund jet plane. This would make the interpretation of the substructure in terms of QCD cascades more challenging, which is the issue addressed via the partial \PQb hadron reconstruction. Therefore, the jet substructure observables are also computed using only charged-particle tracks. Using only the charged particles has the additional advantage of yielding better angular and momentum resolution. A ${\pttrack>1\GeV}$ selection is applied in the CA reclustering step, in order to avoid the fast decrease in track reconstruction efficiency and minimize possible data-to-simulation differences. Although the charged-particle jet substructure is not collinear safe, comparisons to theoretical predictions of certain observables may still be possible by making additional nonperturbative corrections to account for the use of only the charged component of the jet. The charged-particle jet substructure variable distributions can be used for MC event generator tuning of the QCD cascade of \PQb jets.

\subsection{The jet fragmentation function}

With a partially clustered \PQb hadron, it is possible to study other aspects of \PQb jet fragmentation, such as the fraction of the jet transverse momentum carried by the \PQb hadron. Since the \PQb hadron is clustered using only charged particles, we compare its \pt to the \pt of the charged-particle components of the jet, 
\begin{equation}
\label{eq:fragmentationFunction}
    \zch = \frac{\ptbch}{\ptjetch},
\end{equation}
where \ptbch is the \pt of the partially reconstructed \PQb hadron and \ptjetch is the \pt of the charged-particle component of the jet. We refer to Eq.~(\ref{eq:fragmentationFunction}) as the \PQb jet fragmentation function. The $\kt>1\GeV$ selection that is used for the groomed jet substructure observables is not used for \zch, so \zch includes information complementary to the one contained in \rg and \zg. There are existing measurements of the fragmentation function of \PQb hadrons using exclusive decays, \eg, $\PBpm \to \JPsi \PKpm$, in jets \cite{ATLAS:2021agf, LHCb:2017llq}. The ATLAS Collaboration has followed a cut-based approach to cluster the \PQb hadron decay daughters for b jets produced in top quark pair production~\cite{ATLAS:2022miz}. In that measurement, the four-momentum of the reconstructed secondary vertex in the event is treated as a proxy for the \PQb hadron in order to measure its fragmentation function and related properties. In the approach presented in this paper, we optimize the partial reconstruction of \PQb hadron candidates using all jet charged-particle constituents and the secondary vertices they are associated with as input to a multivariate analysis. Charged particles not associated with a secondary vertex are also considered.

\section{Analysis method and corrections}\label{sec:method}

\subsection{The \texorpdfstring{\PQb}{b} tagging efficiency scale factors}
A subset of \PQb jets is selected from the inclusive jet sample, using the \textsc{ParticleNet} tagger~\cite{Qu:2019gqs}.
The resulting signal efficiency in simulation may differ from the one found in data depending on the jet properties.
In this analysis, we only measure per-jet normalized quantities, which are insensitive to an overall data-to-simulation difference in tagging performance.  
However, data-to-simulation differences as functions of the observables are potential sources of systematic uncertainty. 
To account for such effects, we evaluate scale factors for the \PQb tagging performance difference of \textsc{ParticleNet} between data and simulation, using the ``reference lifetime algorithm'' described in Ref.~\cite{CMS:2012feb}.
The scale factors are applied as a bin-by-bin correction on the detector-level distributions of \PQb jets, in order to mimic the \PQb tagging performance in the MC simulation used in the unfolding procedure.

\subsection{Residual background subtraction}
The set of \PQb-tagged jets in the considered kinematic range consists of 83\% jets with exactly one \PQb hadron (single-\PQb jets), 16\% jets with two or more \PQb hadrons (double-\PQb jets) and 1\% jets coming from gluons, and light and \PQc quarks (light+\PQc jets), according to the simulated sample from \PYTHIA{8} CP5.
The main residual background contamination coming from double-\PQb jets, resulting for example from gluon splitting processes, is detrimental to the measurement, as the two or more \PQb hadrons are merged into one during the partial reconstruction process.
This effect is minimized by fitting the partially reconstructed \PQb hadron mass distribution to templates coming from simulation.
The template fit is performed in bins of the measured observables. 

\begin{figure}[ht]
    \centering
    \includegraphics[width=0.325\textwidth]{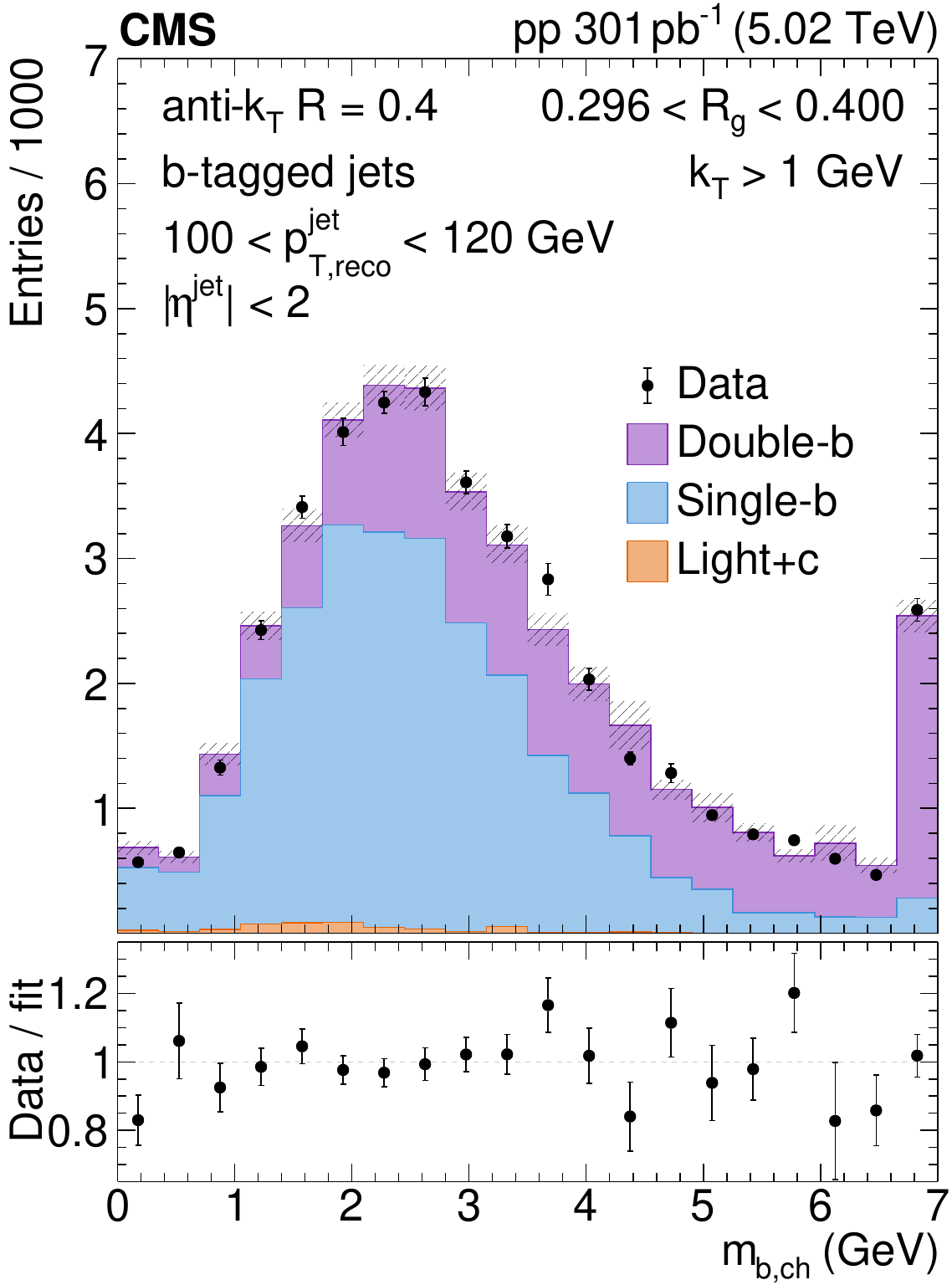}
    \includegraphics[width=0.325\textwidth]{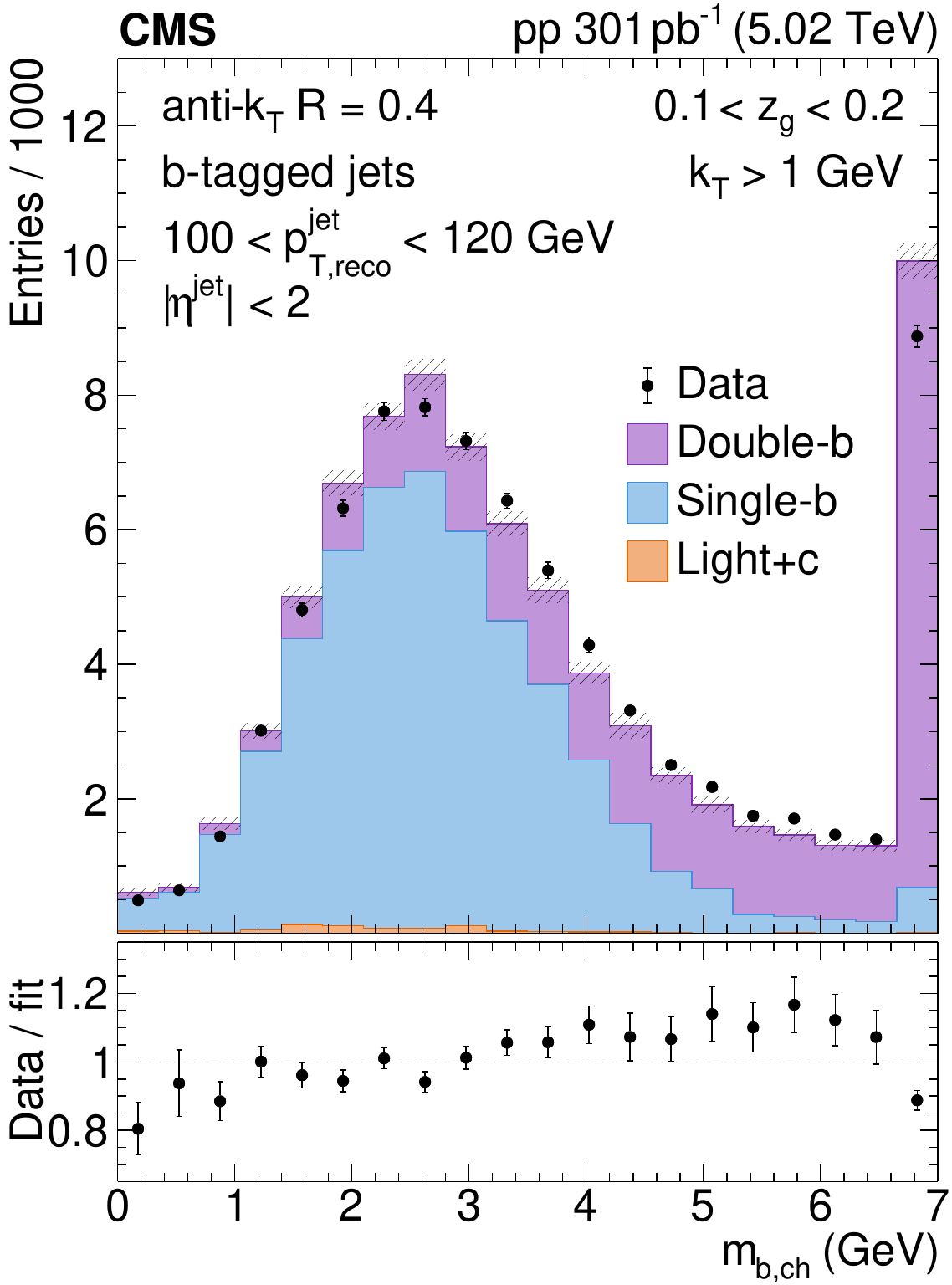}
    \includegraphics[width=0.325\textwidth]{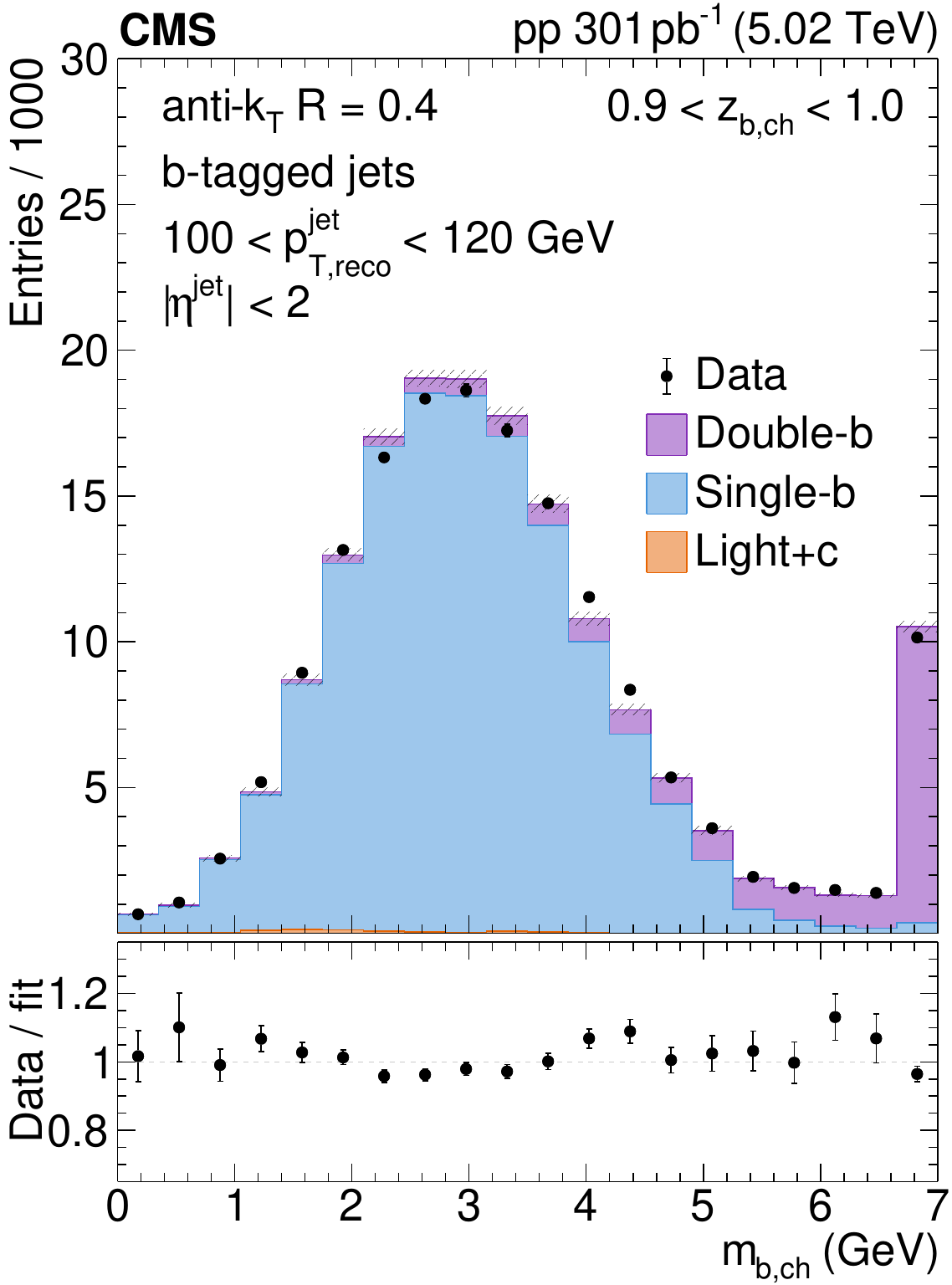}
    \caption{Examples of the fit of the partially reconstructed \PQb hadron mass to double-\PQb, single-\PQb, and light+\PQc jet templates coming from \PYTHIA{8} CP5. From left to right, the distributions in selected \rg, \zg, and \zch bins are presented. The black points represent the data counts, while the violet, blue, and orange filled curves represent the double-\PQb, single-\PQb, and light+\PQc jet templates, with the fractions acquired from the fit. The ratio between the data and the fit values is presented in the lower panels.}
    \label{fig:template_fit}
\end{figure}

We use three templates: one for single-\PQb jets, one for double-\PQb jets, and one light+\PQc jets.
Because of the high \PQb jet purity of the selected \PQb tagging working point, the light+\PQc background is strongly suppressed and the corresponding template is poorly populated. 
It is therefore combined with the single-\PQb template, fixing the light+\PQc over single-\PQb fraction to the value obtained from the \PYTHIA{8} CP5 simulation. We assign an uncertainty associated with this template combination step, as described in Section \ref{sec:syst_uncertainties}. Additionally, we assign an uncertainty associated with the template shape by using \HERWIG{7} CH3 as an alternative MC generator, which is also discussed in Section \ref{sec:syst_uncertainties}. Examples of the template fit performed in one bin of each observable are given in Fig.~\ref{fig:template_fit}.
The detector-level yield distributions, with the aforementioned \PQb tagging efficiency scale factors applied, are multiplied by the extracted single-\PQb jet fractions determined from the template fitting procedure.

\subsection{Unfolding}
Bin-to-bin migration effects are corrected for using two-dimensional unregularized unfolding with the \textsc{RooUnfold} package \cite{Brenner:2019lmf, Adye:2011gm}, in bins of jet \pt and the observable of interest. 
The jet \pt includes three bins, where the middle bin is the one reported in the measurement, and the bins below and above account for the bulk of the migrations.
The fragmentation function \zch spans the whole range from 0 to 1, where the edges are included in the first and last bins, respectively. 
Regarding the groomed observables, the jets that fail the SD condition or have $\kt<1\GeV$ are included in a dedicated bin, such that they can migrate from SD-untagged to SD-tagged from the particle to the detector level or vice versa.
In the case of \rg, an overflow bin of large $\ln(R/\rg)$ (small \rg) is also included.
This ensures that all possible migrations are taken into account, and the same collection of jets is considered for all three observables.
Both the detector and particle levels correspond to the charged-particle substructure and use the same binning.

The fraction of reconstructed jets without a generated jet counterpart is found to be negligible. 
A reconstruction purity bin-by-bin correction is applied, to account for reconstructed jets which were generated with a \pt outside the kinematic range. 
In the reported jet \pt range of 100 to 120\GeV, this corresponds to 3--7\% of reconstructed jets in the \PYTHIA{8} CP5 samples.  
The resulting distributions are then unfolded to the charged-particle level using unregularized matrix inversion. 
The fraction of jets migrating to neighboring bins is about 50\%, with comparable migration rates in both the kinematic and substructure dimensions. 
The charged-particle-level result is corrected for the reconstruction efficiency, which accounts for generated jets that were reconstructed outside the jet \pt range. This corresponds to a 3--6\% correction in each bin. 
 
\subsection{The \texorpdfstring{\PQb}{b} tagging efficiency correction}
The charged-particle-level result is corrected for biases that are introduced as a result of the use of \PQb tagging. 
All \PQb tagging algorithms are prone to selecting \PQb jets with specific qualities which make them easier to identify, such as high-multiplicity decay vertices.
To account for this bias, we calculate a \PQb tagging efficiency correction at the particle level of the simulation.
This correction corresponds to the inverse of the \PQb jet efficiency in each bin and is applied as a multiplicative per-bin factor, as a function of the jet substructure observable. 

A sample of inclusive jets is used as a reference for the substructure variables. These serve as a proxy for a sample of jets where mass effects are absent, due to the dominance of light-quark and gluon jets in inclusive jets. The inclusive jet distributions are also unfolded to the stable-particle level. 

\section{Systematic uncertainties}\label{sec:syst_uncertainties}

Systematic uncertainties originating from several theoretical and experimental sources are considered in this study. 
The individual variations are propagated through the relevant steps in the analysis and the fully corrected result is compared to the nominal one. 
All relative uncertainties resulting from a single variation, such as the shower and hadronization model, are symmetrized around zero. 
In two-point variations, \ie, varying a parameter up and down within its uncertainty, such as the jet energy resolution, the larger of the two resulting uncertainties is chosen and symmetrized in each bin. 
The following systematic uncertainties apply both to inclusive and \PQb jets and are propagated through the corrections to the stable-particle level.

\textit{Shower and hadronization modeling uncertainty:} 
This uncertainty is estimated by building the response matrix, along with the reconstruction purity and efficiency corrections, using events simulated by the \HERWIG{7} CH3 generator instead of the nominal \PYTHIA{8} CP5. Since \HERWIG{7} CH3 has a different parton shower algorithm and a different hadronization model, this uncertainty mostly quantifies the difference in the detector response at the full-jet and the subjet levels. 

\textit{Parton shower scale uncertainties:}
To estimate the theoretical uncertainty due to missing higher-order corrections in the perturbative calculation of the parton shower of \PYTHIA{8} CP5, we consider variations of the renormalization scale for the final-state radiation (FSR) and initial-state radiation (ISR) by factors of 2 and 1/2, corresponding to up and down variations, independently for the two sources~\cite{Cacciari:2003fi, Catani:2003zt}. 
In practice, the nominal \PYTHIA{8} CP5 generated sample is reweighted at the particle level as a function of the jet \pt and of the substructure variable of interest in order to reproduce the distribution resulting from the FSR and ISR variations. The unfolding is repeated for each of these variations.
Generally, these up and down variations have symmetric effects on the jet substructure variable distributions and in the associated systematic uncertainties. The effect of the variations of the ISR renormalization scale are smaller than the ones from FSR.
The resulting uncertainties are symmetrized per source (FSR, ISR) and added in quadrature.

\textit{Tracking efficiency uncertainty:} 
Track reconstruction in high-density environments, such as the core of a jet, is prone to mismodeling in simulation.
To account for that, we propagate a tracking reconstruction efficiency uncertainty through the unfolding procedure.
This is done by randomly removing 3\% of the reconstructed tracks in simulation during the substructure extraction.
This value covers the residual differences observed in data and simulation for tracking in the jet core in the context of jet energy scale determination~\cite{CMS:2016lmd}.

\textit{Response matrix statistical uncertainties:} 
We assign a systematic uncertainty to the measurement due to the finite number of simulated events that populate the response matrix by employing the ``jackknife'' resampling technique~\cite{5685f092-0b53-33cc-a858-e3a8f62b8477,6c956df0-ca97-3419-9961-dcc097853946,640d49bc-15a9-3895-95e3-0be4499a5fe8}, as used previously in Ref.~\cite{CMS:2018vzn}. In this method, we create ten samples, each with 90\% of the simulated events, so that each event is excluded exactly once in the resampling procedure. We then obtain ten different fully corrected distributions, using each of these samples in the unfolding. The standard deviation of this ensemble of distributions, scaled by 10/9 to account for using 90\% of the full sample for each subset of events, is taken as the statistical uncertainty from the limited sample size of the MC simulation.

\textit{Jet energy scale and resolution uncertainties:} 
The jet energy scale uncertainty is estimated by varying the detector-level jet \pt in simulation within the $\eta$-\pt dependent jet energy scale uncertainty~\cite{CMS:2016lmd}. It is decomposed into multiple sources, and the unfolding is repeated for each subcomponent. The dominant uncertainty for the inclusive jets comes from the flavor component, as a result of the mixture of quark and gluon jets, while it is smaller for the \PQb jets. The per-source symmetrized uncertainties are added in quadrature. The jet energy resolution uncertainty is also considered. As described in Section \ref{sec:cms_detector}, jet energy resolution scale factors are already applied to the detector-level jet \pt in simulation in order to account for differences between simulation and data. An uncertainty in the measurement is derived by varying the aforementioned scale factors within their uncertainties. 

\begin{figure}[ht]
    \centering
    \includegraphics[width=0.99\textwidth]{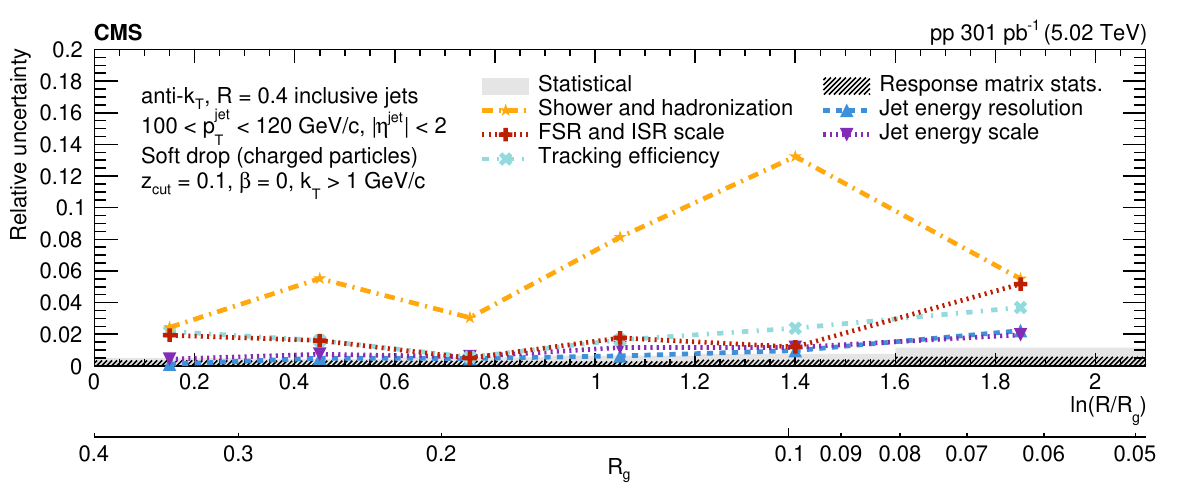}
    \includegraphics[width=0.99\textwidth]{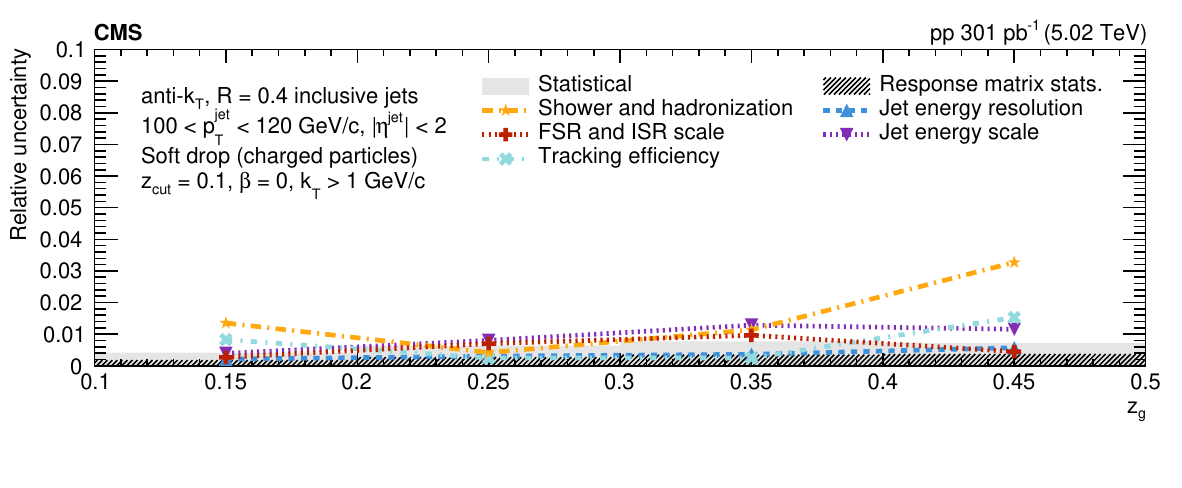}
   \caption{The breakdown of systematic uncertainties for inclusive jets (upper panel: \rg, lower panel: \zg). The statistical uncertainty is also shown in gray shaded boxes.}
   \label{fig:syst_incl}
\end{figure}

There are also uncertainties considered that only apply to \PQb jets. These uncertainties relate to the \PQb tagging and background subtraction procedures and are applied to the corresponding steps of the analysis.

\textit{\PQb tagging efficiency uncertainty:}
As described in Section \ref{sec:method}, an efficiency correction is applied to account for the bias introduced by the use of \PQb tagging.
This correction is calculated from the \PYTHIA{8} CP5 simulation.
To account for differences in the \PQb tagging performance between the simulation and the data, scale factors are applied to the data prior to the unfolding.
We assign an uncertainty in the \PQb tagging efficiency by varying the scale factors within their uncertainty. 

\textit{Residual background subtraction uncertainties:}
We consider two sources of systematic uncertainty related to the method used for the subtraction of the residual gluon, light-quark, and \PQc quark jet components, as well as jets including more than one \PQb quark, from the \PQb-tagged jet collection.
In one case, we replace the nominal templates based on \PYTHIA{8} CP5 simulated events with the ones derived from the \HERWIG{7} CH3 generator, which affects the template shapes.
The estimated fraction of \PQb jets in the data sample is found to be higher when using the \HERWIG{7} CH3 templates.
In the other case, we vary the MC fraction of light+\PQc relative to the single-\PQb jets in the simulation. This corresponds to an up and down variation of the fraction by 100\% before the two templates are combined, as described in Section \ref{sec:method}.
In both cases, a new single-\PQb jet fraction is extracted and propagated. 
We then compare these variations with the nominal distributions and extract the \PQb jet fraction and light+\PQc misidentification rate uncertainties, respectively. Each of these pieces is considered to be independent from each other.

\begin{figure}[!ht]
\centering
\includegraphics[width=0.99\textwidth]{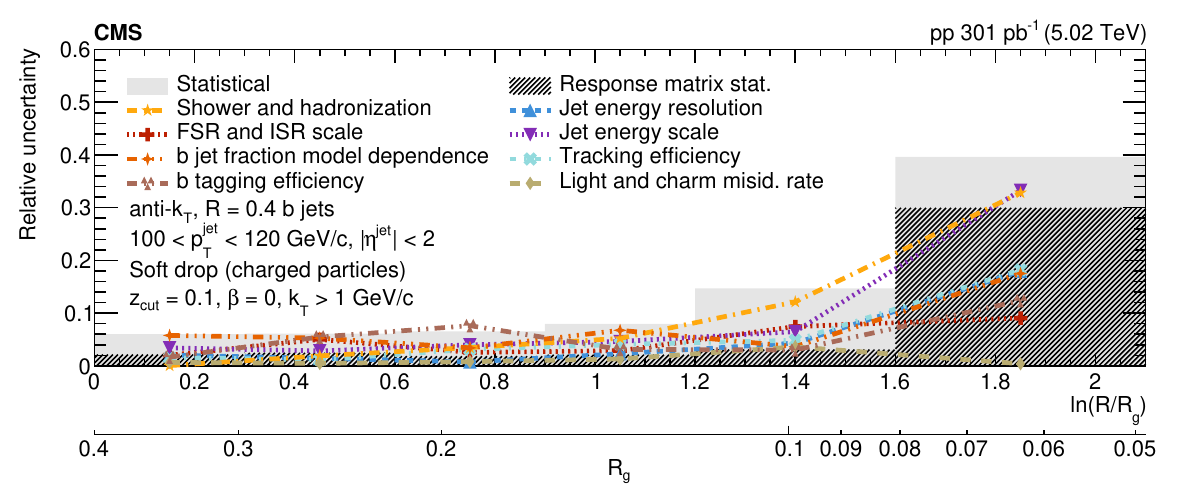}
\includegraphics[width=0.99\textwidth]{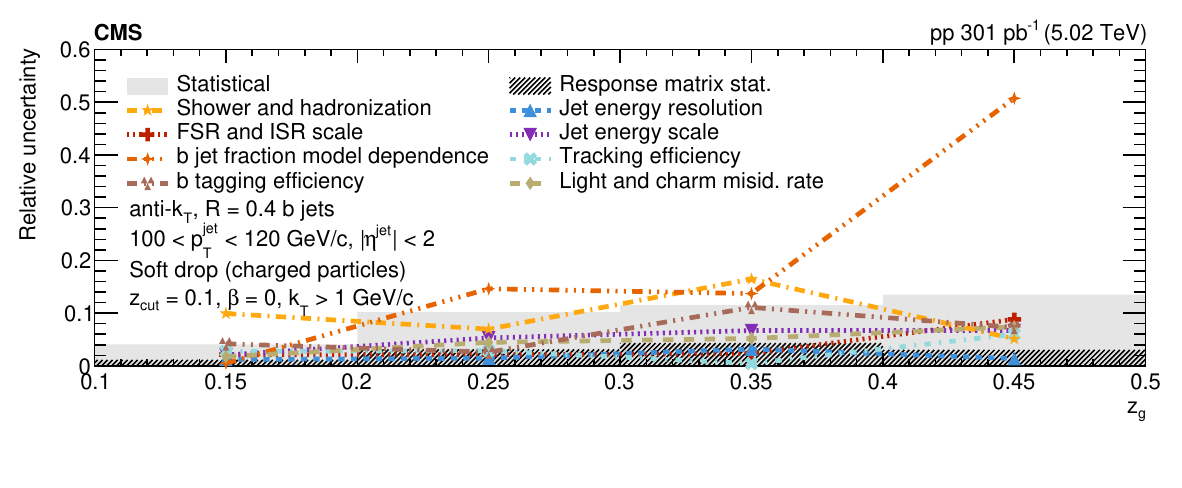}
\includegraphics[width=0.99\textwidth]{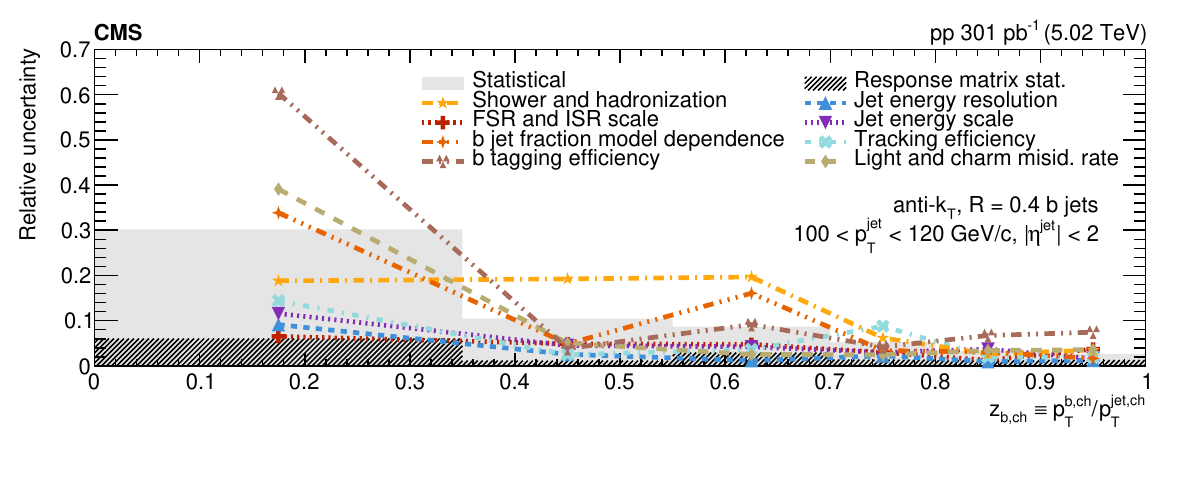}
\caption{The breakdown of systematic uncertainties for \PQb jets (upper panel: \rg, middle panel: \zg, lower panel: \zch). The statistical uncertainty is also shown in gray shaded boxes.}
\label{fig:syst_btag}
\end{figure}

The size of the systematic uncertainties is presented in Fig.~\ref{fig:syst_incl} for inclusive jets and Fig.~\ref{fig:syst_btag} for single-\PQb jets.
The leading uncertainties for both samples of jets are related to the physics model, which affects the detector response at the substructure and full jet levels. Namely, the shower and hadronization model, the parton shower scale and the model dependence of the \PQb jet fraction are dominant, reaching up to 14\% uncertainty for inclusive jets and 20\% for \PQb jets, with the exception of the smallest \rg (largest $\ln(R/\rg)$) bin. An improved description of the MC generators could improve these uncertainties in the future.
In the case of \PQb jets, we also observe the \PQb tagging efficiency uncertainty become important, especially in the small \zch region.

\section{Results}\label{sec:results}

\begin{figure}[th]
     \centering
     \includegraphics[width=0.49\textwidth]{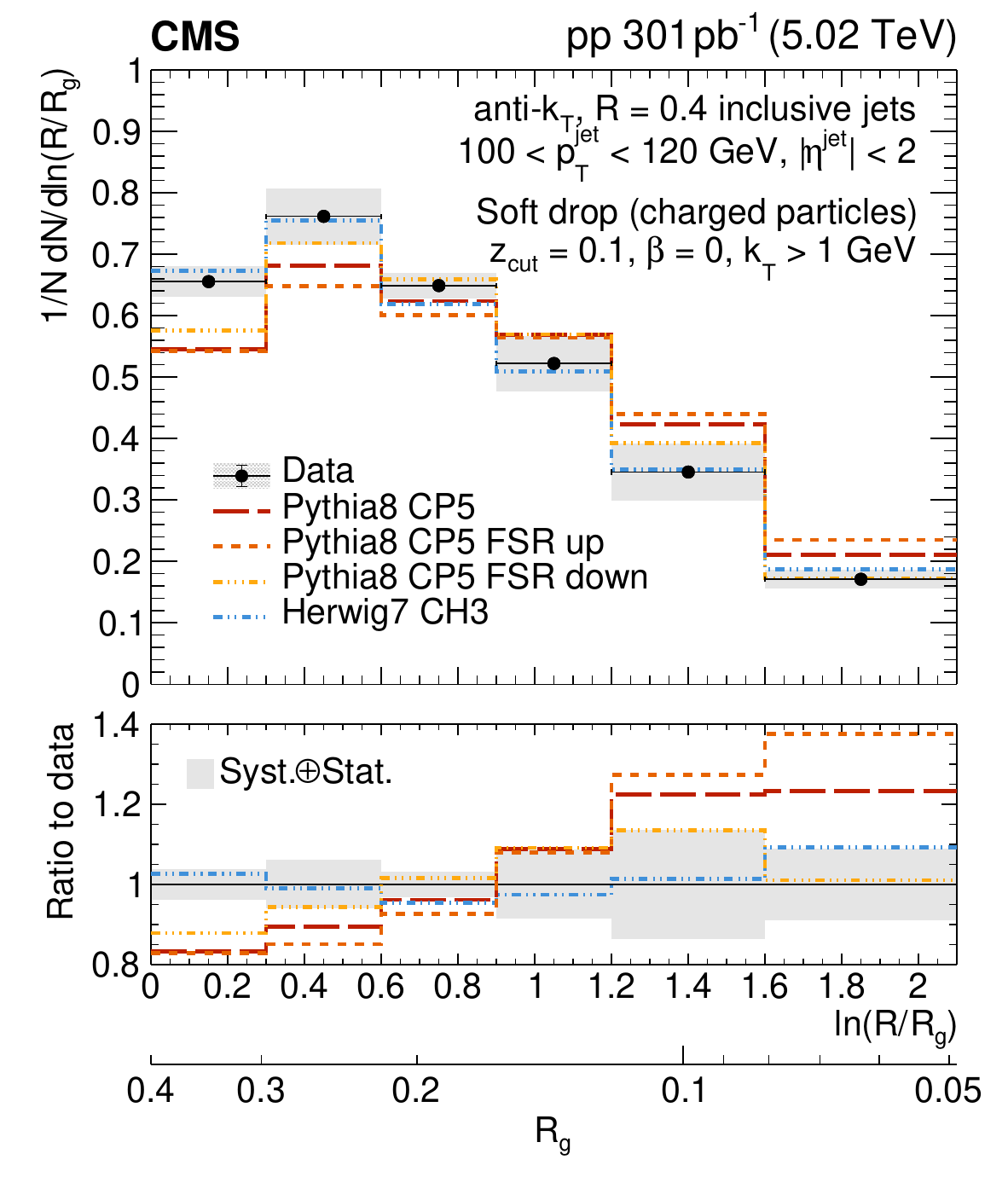}
     \includegraphics[width=0.49\textwidth]{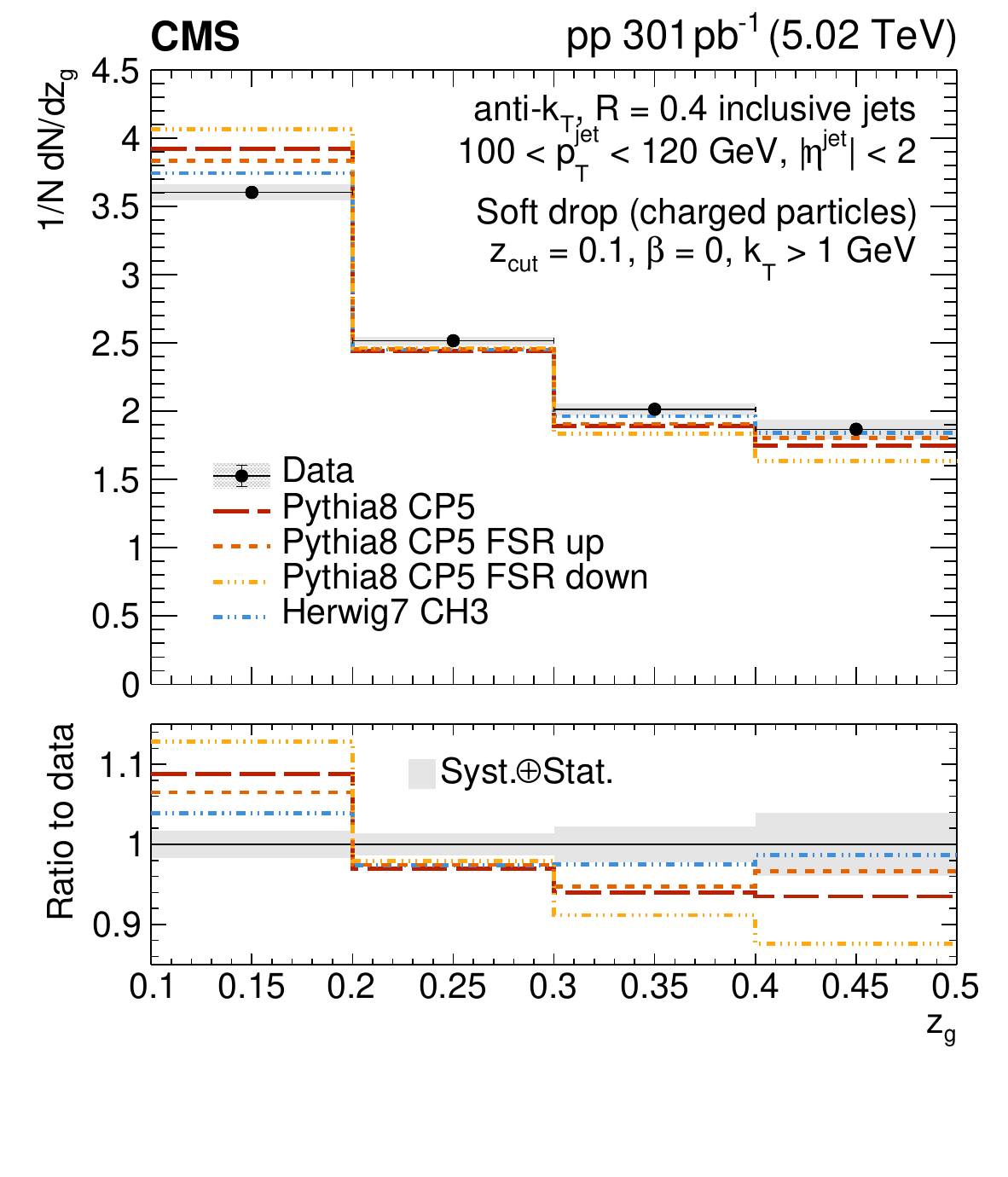}
    \caption{Distributions of groomed substructure observables \rg (left) and \zg (right) corrected to the charged-particle level for inclusive jets. The gray band represents the systematic uncertainties added in quadrature, while the vertical bars represent the statistical uncertainty. Distributions coming from the \PYTHIA{8} CP5 and \HERWIG{7} CH3 MC event generators are also presented. The ratio of the MC simulations to the data is presented in the lower panels.}
    \label{fig:data_vs_mc_incl}
\end{figure}

\begin{figure}[ht]
     \centering
     \includegraphics[width=0.49\textwidth]{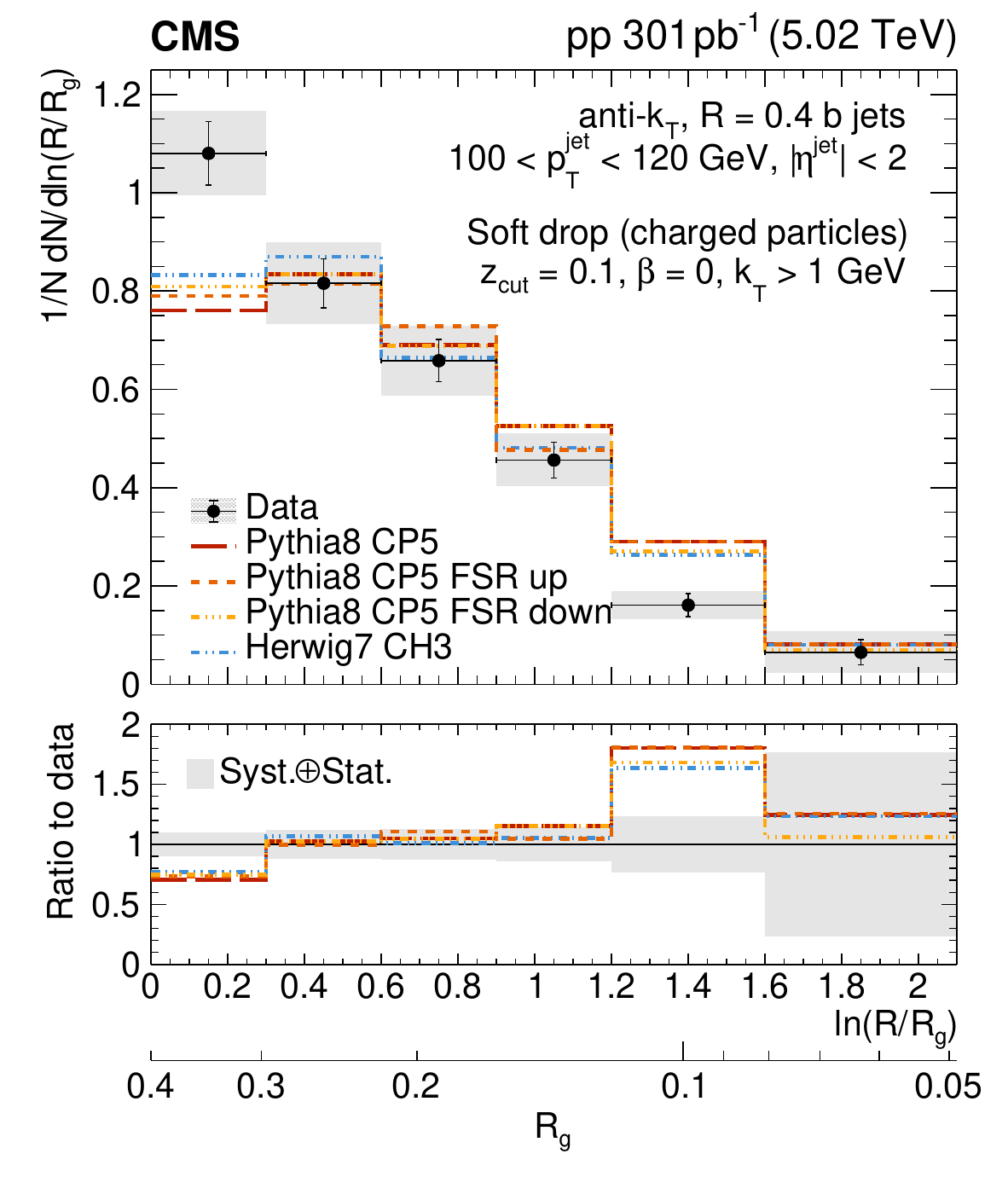}
     \includegraphics[width=0.49\textwidth]{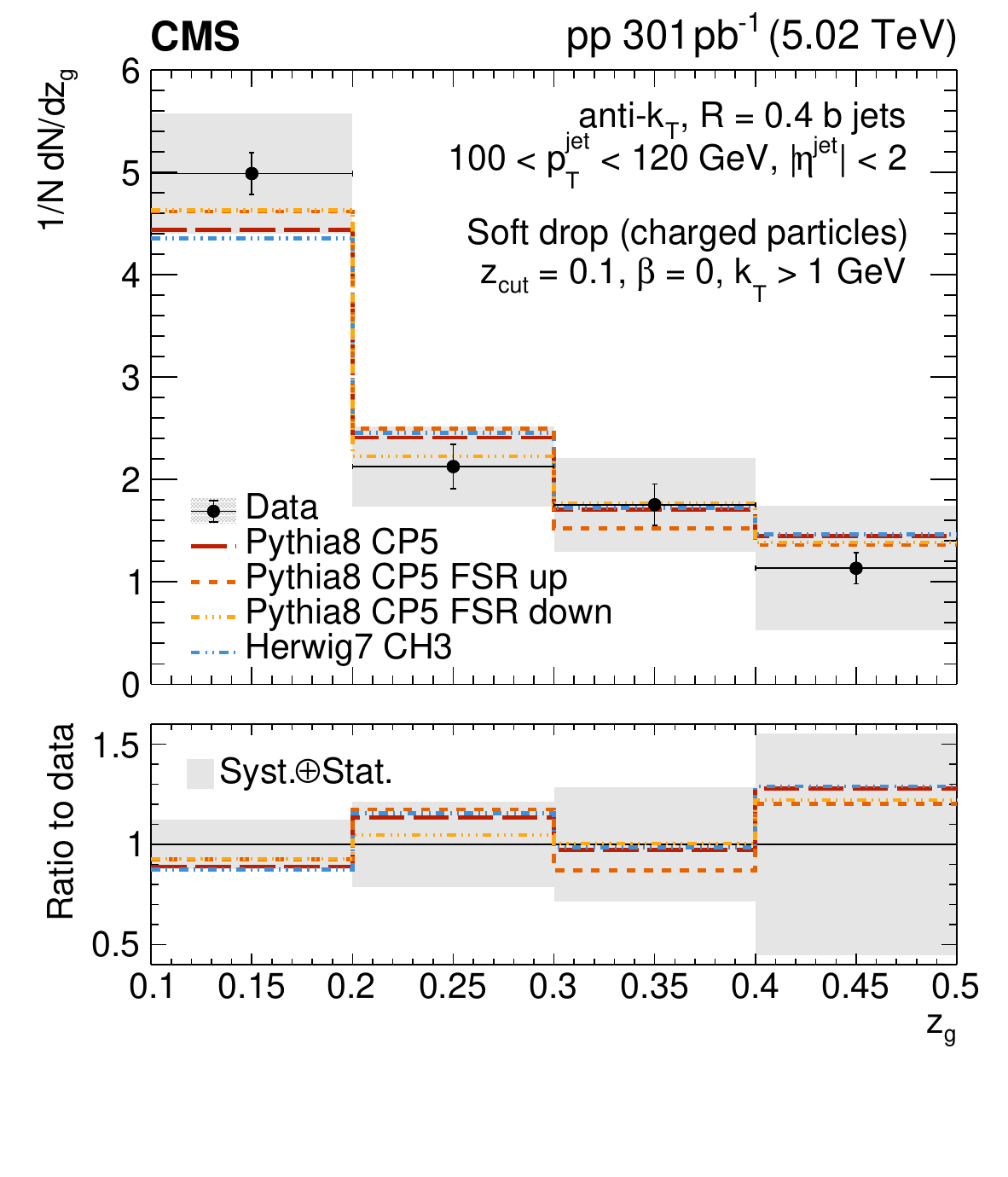}
    \caption{Distributions of the groomed substructure observables \rg (left) and \zg (right) corrected to the charged-particle level for \PQb jets. Distributions coming from the \PYTHIA{8} CP5 and \HERWIG{7} CH3 MC event generators are also presented. The ratio of the MC simulations to the data is presented in the lower panels.}
    \label{fig:data_vs_mc_btag}
\end{figure}

The fully corrected distributions of \rg and \zg for inclusive jets and \PQb jets are presented in Figs.~\ref{fig:data_vs_mc_incl}~and~\ref{fig:data_vs_mc_btag}, respectively. 
The fraction of jets that do not satisfy the SD grooming condition or have ${\kt<1\GeV}$ is found to be ${38.5\pm 0.1\stat\pm 2.3\syst\%}$ for inclusive jets, whereas \PYTHIA{8} CP5 (\HERWIG{7} CH3) predicts 40.7\% (36.3\%).
For \PQb jets, the fraction is found to be ${66.2\pm 0.9\stat\pm 2.7\syst\%}$ in the data and 71.0\% (62.3\%) in \PYTHIA{8} CP5 (\HERWIG{7} CH3).

\begin{figure}[ht]
     \centering
     \includegraphics[width=0.49\textwidth]{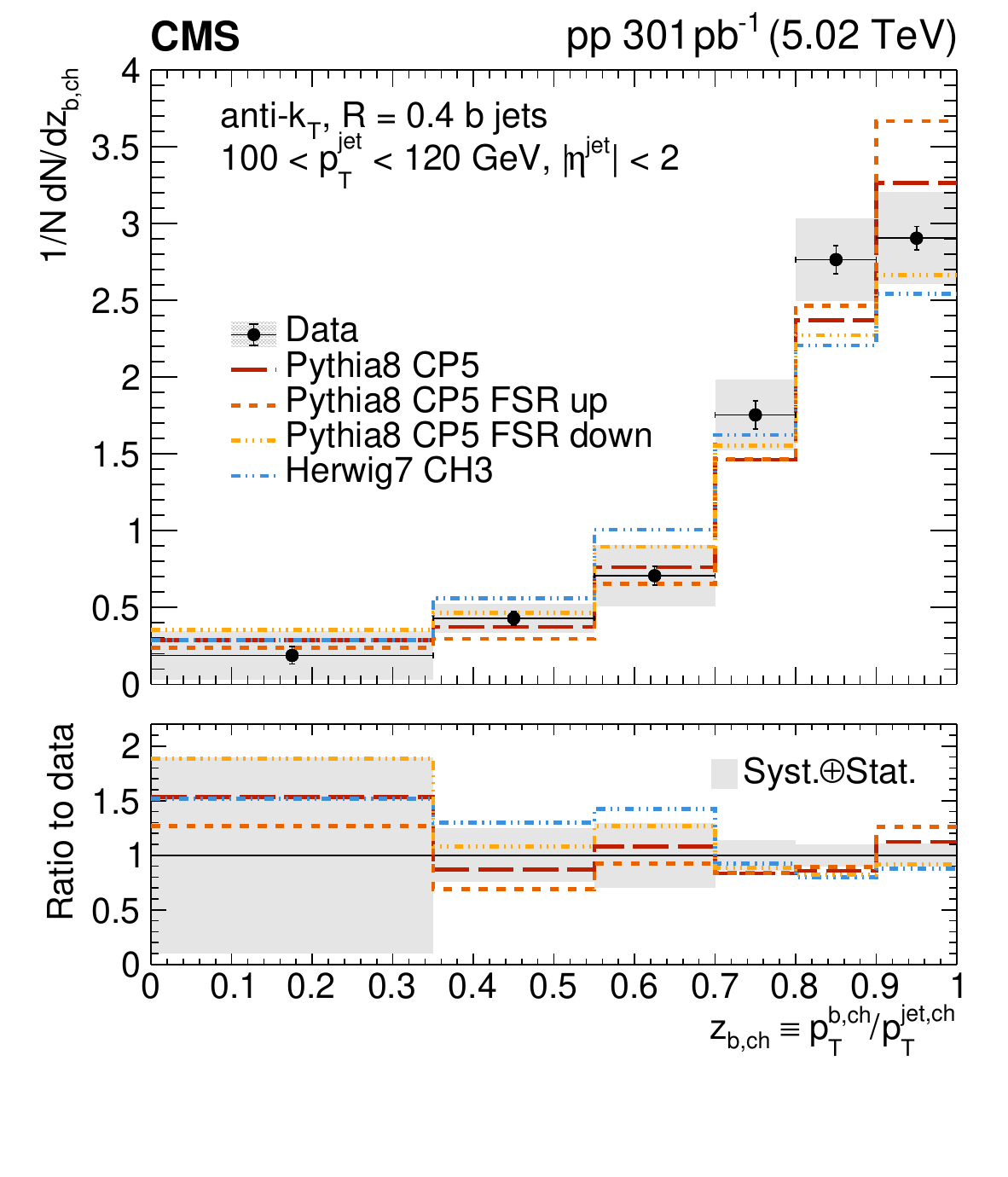}
    \caption{The distribution of the jet fragmentation function \zch corrected to the charged-particle level. Distributions coming from the \PYTHIA{8} CP5 and \HERWIG{7} CH3 MC event generators are also presented. The ratio of the MC simulations to the data is presented in the lower panel.}
    \label{fig:data_vs_mc_btag_z}
\end{figure}

Strong deviations are observed from the \PYTHIA{8} CP5 distributions for inclusive jets. 
More specifically, the \PYTHIA{8} CP5 \rg distribution is shifted to larger $\ln{(R/\rg)}$ (smaller \rg) values with respect to the data. 
The \PYTHIA{8} CP5 prediction also slightly overestimates the number of jets with a small momentum balance \zg between the two subjets. 
On the other hand, \HERWIG{7} CH3 has a good agreement with the data in both observables.

In the case of \PQb jets, deviations are also observed for \rg, while the two generators agree with the data within the experimental uncertainties for \zg.
The agreement of \HERWIG{7} CH3 with the measured \rg distribution is less good for \PQb jets than for inclusive jets, suggesting that there might be mismodeling as a result of the presence of mass effects.

The jet fragmentation function \zch, \ie, the charged-particle momentum fraction of the partially reconstructed \PQb hadron with respect to the jet, and its comparison with the simulated predictions, is presented in Fig.~\ref{fig:data_vs_mc_btag_z}.
The distribution peaks at high \zch, indicating that the \PQb hadron holds most of the momentum during the fragmentation process.
The \PYTHIA{8} CP5 prediction agrees with the data within the experimental uncertainty, while the \HERWIG{7} distribution has a slightly larger width.

\begin{figure}[ht]
     \centering
     \includegraphics[width=0.49\textwidth]{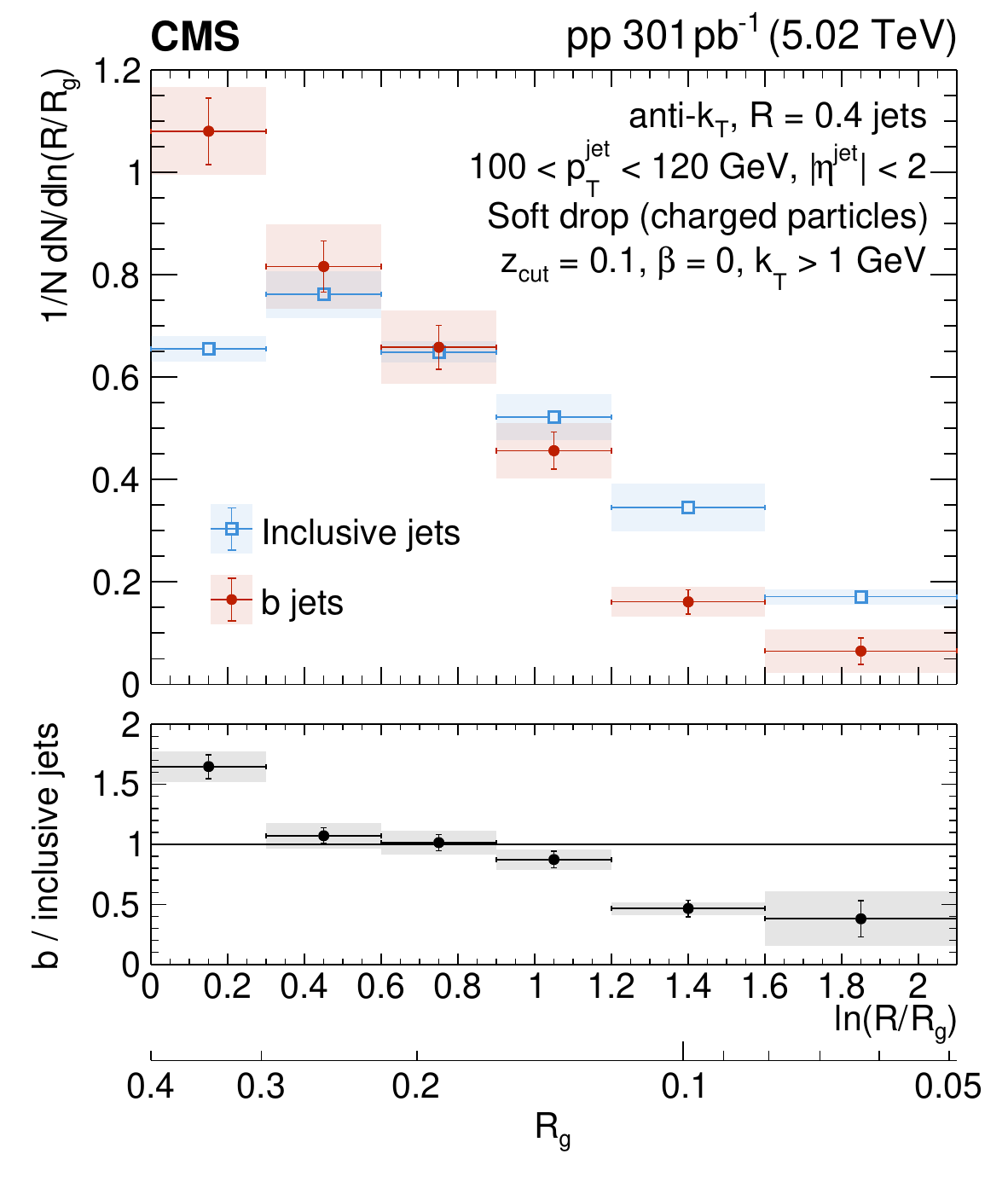}
     \includegraphics[width=0.49\textwidth]{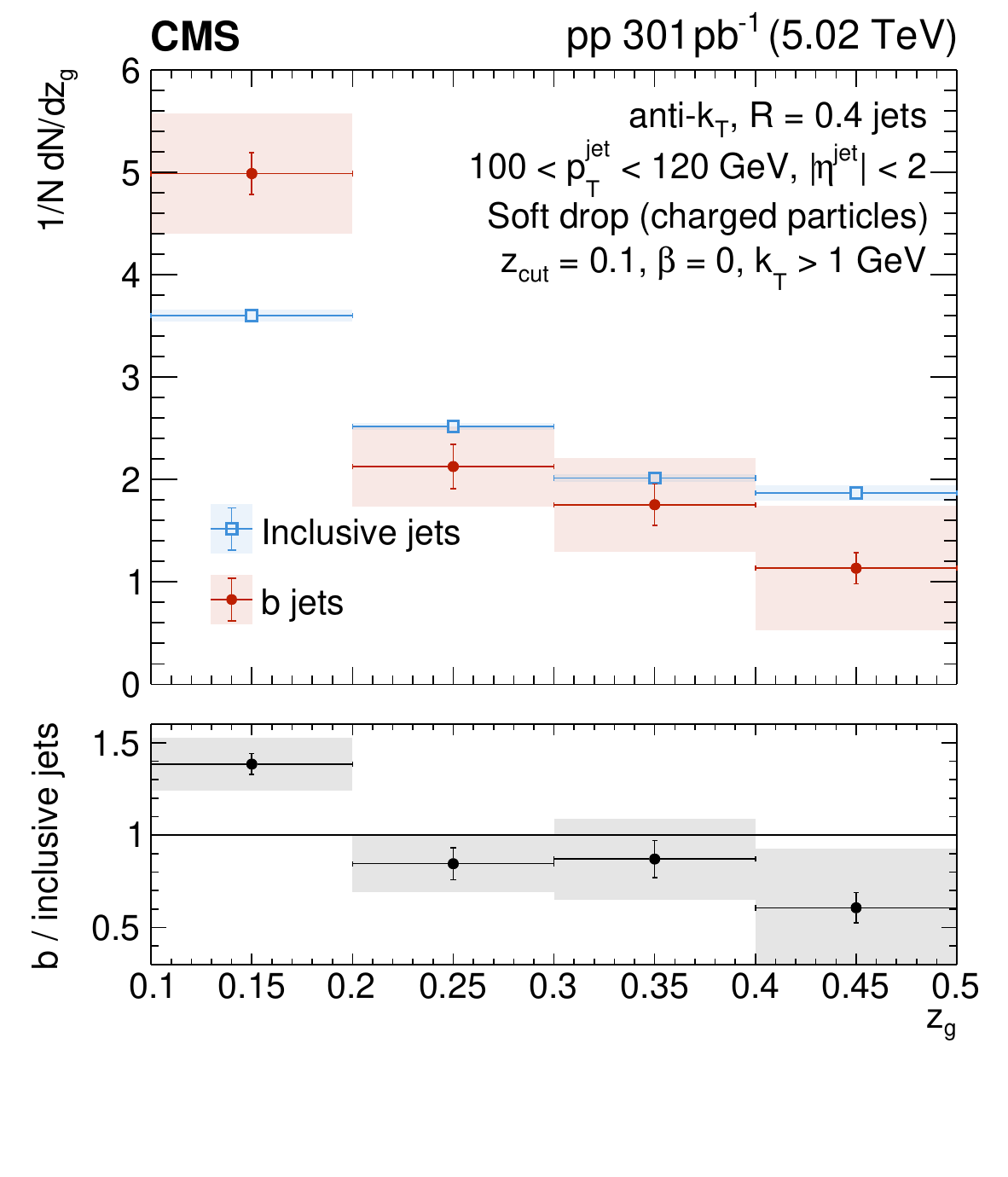}
    \caption{A comparison of the groomed observables \rg (left) and \zg (right) between \PQb and inclusive jets. 
    Most sources of systematic uncertainty are considered fully correlated in the ratio, which is presented in the lower panels, with the exception of those related to flavor and the response matrix. 
    }
    \label{fig:btag_vs_incl}
\end{figure}

\begin{figure}[ht]
     \centering
     \includegraphics[width=0.49\textwidth]{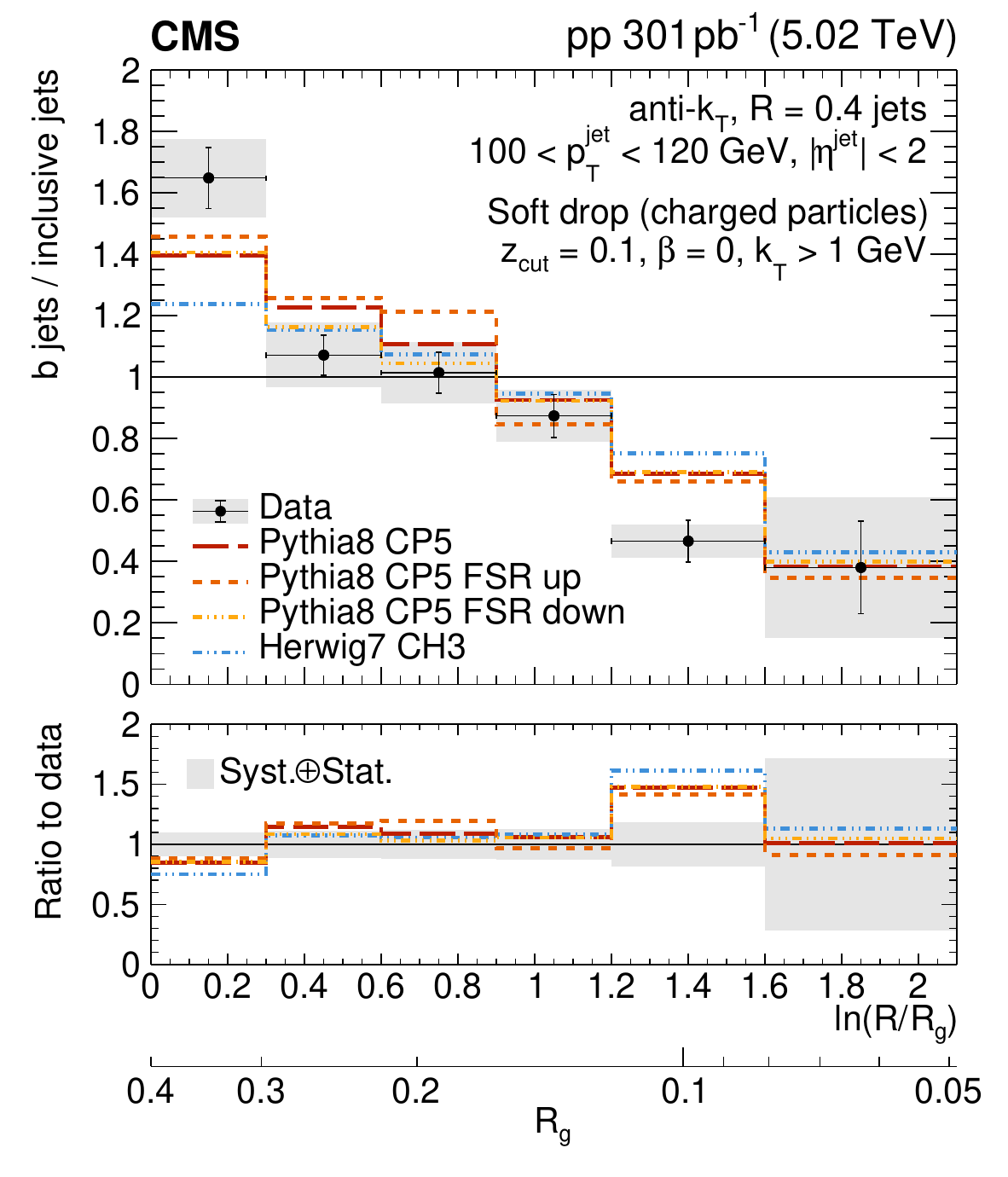}
     \includegraphics[width=0.49\textwidth]{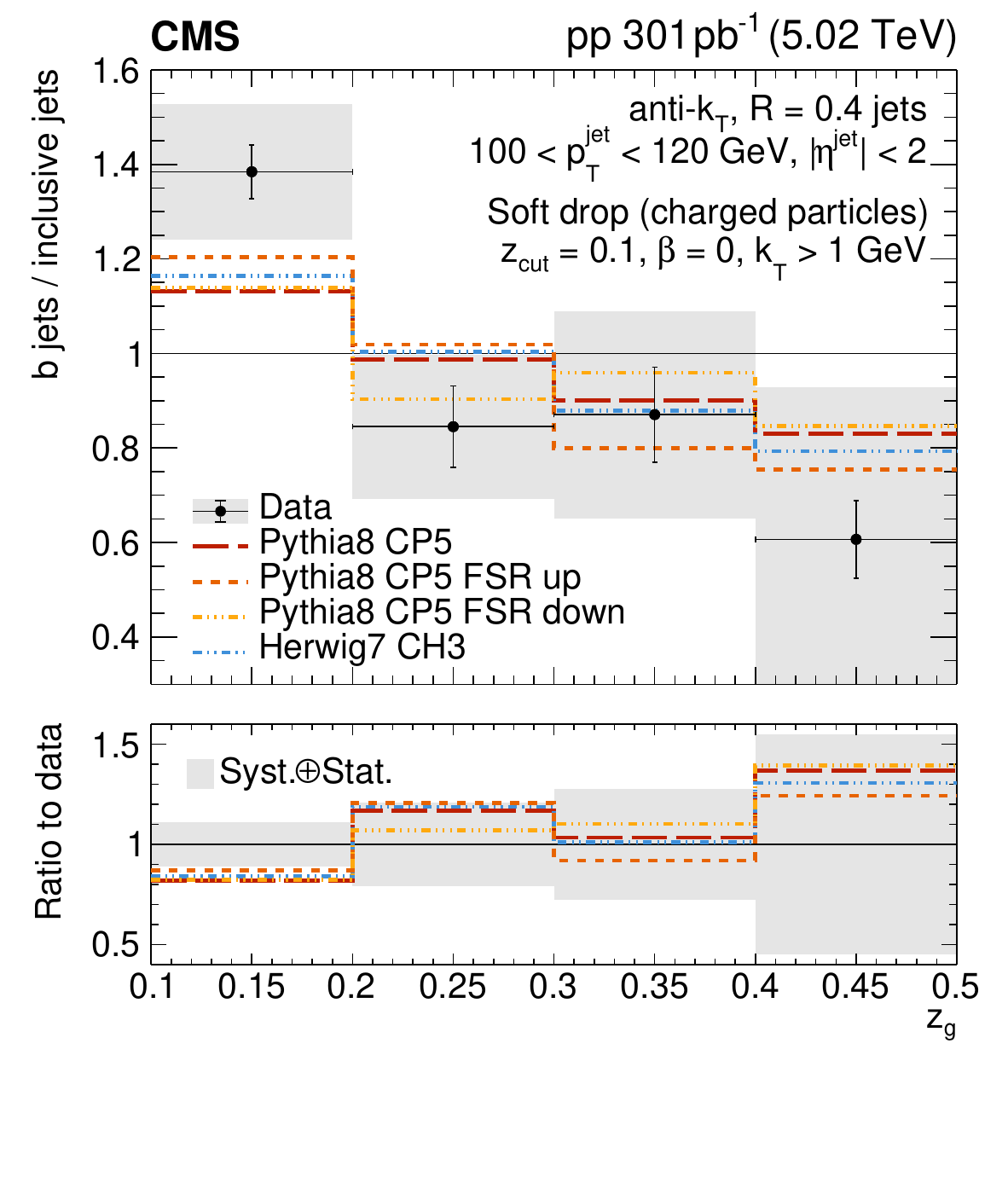}
    \caption{The ratio of the groomed observables \rg (left) and \zg (right) between \PQb and inclusive jets compared to the \PYTHIA{8} CP5 and \HERWIG{7} CH3 MC event generators.
    Most sources of systematic uncertainty are considered fully correlated in the ratio, with the exception of those related to flavor and the response matrix. The ratio of the MC simulations to the data is presented in the lower panels.
    }
    \label{fig:btag_vs_incl_ratio}
\end{figure}

Finally, we present the comparison between inclusive jets and \PQb jets for the groomed observables in Fig.~\ref{fig:btag_vs_incl} and comparisons with the predictions in Fig.~\ref{fig:btag_vs_incl_ratio}. 
Each source of systematic uncertainty is propagated as correlated in the ratio, assuming they are fully correlated bin-by-bin. 
This is with the exception of the response matrix statistical uncertainties, \PQb tagging uncertainties, as well as the uncertainties associated with the background subtraction, which are assumed to be fully uncorrelated. 
In the case of the jet energy scale uncertainties, the flavor-related components are decorrelated for the \PQb jet to inclusive jet ratio. 
The inclusive jet sample is dominated by light-quark and gluon jets, so it serves as a proxy of the substructure of jets in the massless limit. 

In the case of \rg, we observe a strong suppression of emissions at small \rg (large $\ln{(R/\rg)}$) values in \PQb jets with respect to the inclusive jet reference. 
The suppression increases monotonically with decreasing \rg (increasing $\ln{(R/\rg)}$).
This is qualitatively consistent with the dead-cone effect expectations. 
As we are comparing two distributions normalized to their integral, the suppression of emissions at small angles leads to a relative increase of large-angle emissions, and vice versa.
The \PYTHIA{8} CP5 predictions, while they do not describe well the \rg distributions of \PQb jets and inclusive jets separately, they reproduce the \PQb-to-inclusive jet ratio better. 
This suggests that the jet shower is mismodeled in a similar way for \PQb jets and inclusive jets, whereas the small-angle emission suppression due to mass effects is modeled well.
On the other hand, \HERWIG{7} CH3 captures less well the shape of the \rg ratio.

The \PQb jets also tend to favor the smaller momentum balance \zg region more than the inclusive jets. 
This indicates that the \PQb quarks take up a larger fraction of the splitting momentum when emitting gluons than other partons, consistent with the qualitative expectation from the DGLAP massive parton splitting function.
Both generators agree with the data in the \zg ratio, within the experimental uncertainties.

\begin{figure}[ht]
     \centering
     \includegraphics[width=0.49\textwidth]{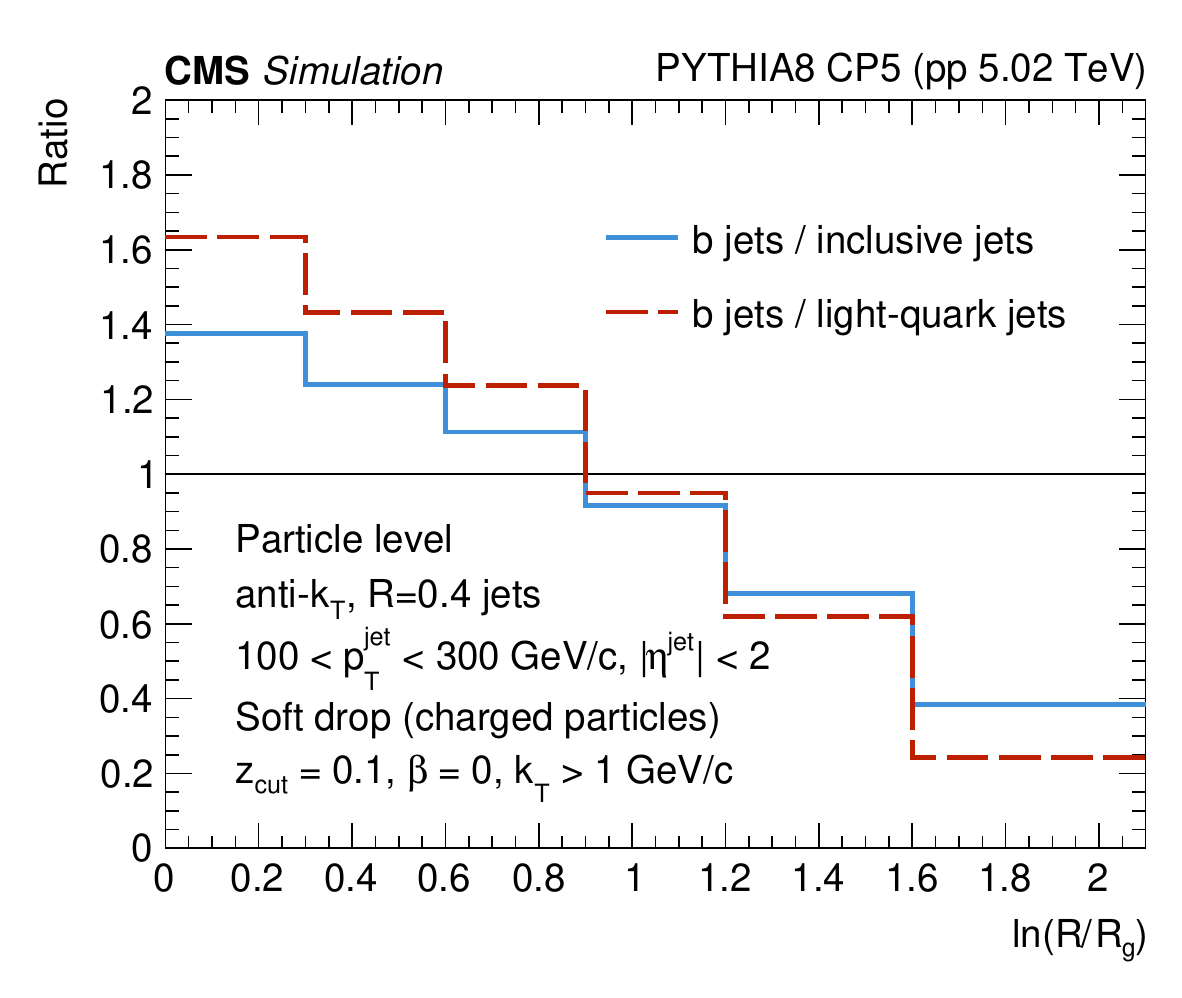}
     \includegraphics[width=0.49\textwidth]{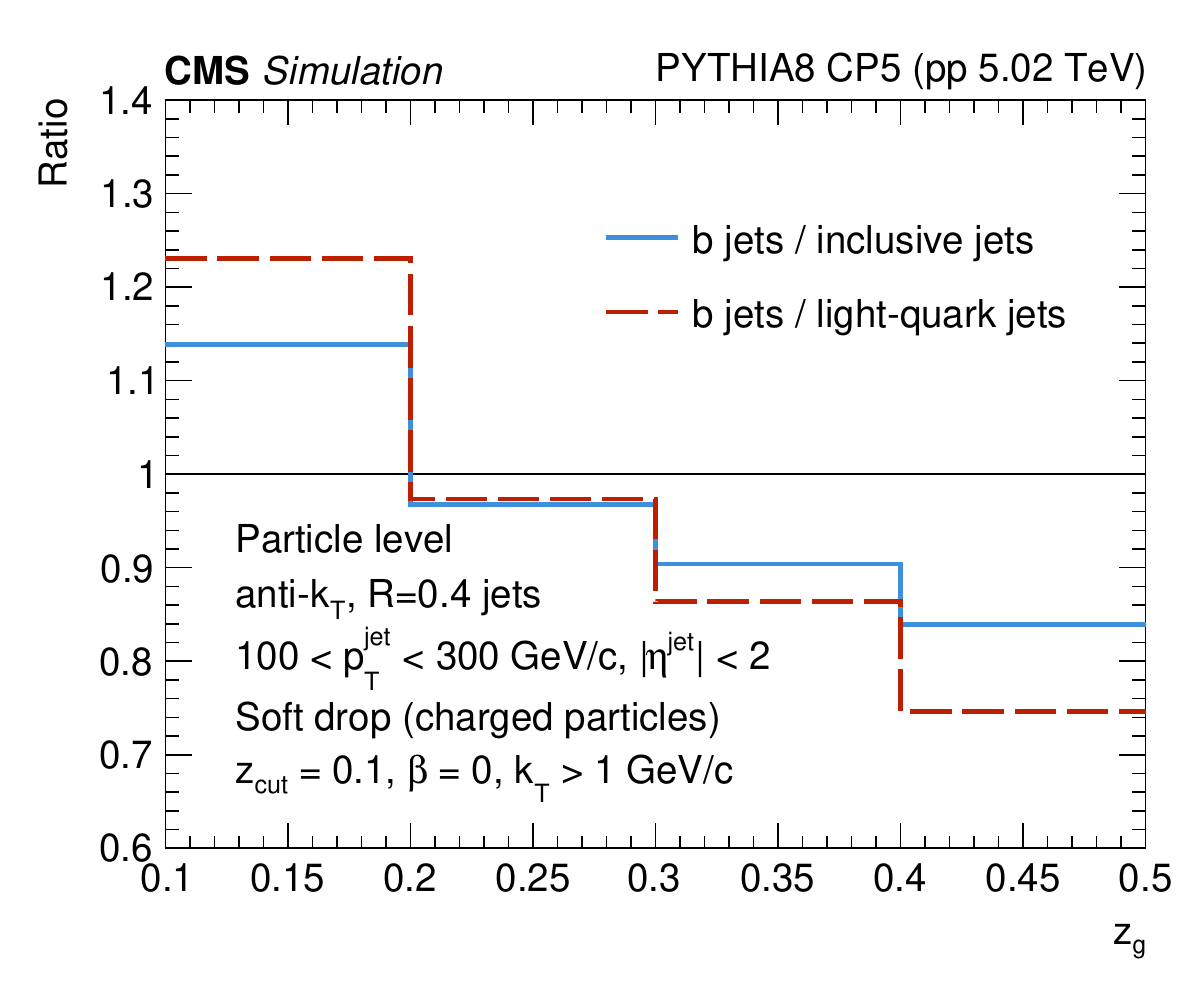}
    \caption{The ratio of the groomed observables \rg (left) and \zg (right) between \PQb and light-quark (dashed red line) or inclusive jets (solid blue line) at the particle level of the \PYTHIA{8} CP5 event generator.
    }
    \label{fig:b_vs_light}
\end{figure}

In this study, the inclusive jet sample is a mixture of quark- and gluon-initiated jets.  
Thus, to illustrate the effect of the heavy-quark mass, it is instructive to compare the ratio of \PQb jet to light-quark jet distributions in \PYTHIA{8} CP5 simulated events, which describe the data well in the \PQb jet over inclusive jet case. 
Figure~\ref{fig:b_vs_light} (left) shows that the suppression of collinear emissions in \PQb jets with respect to light-quark jets is stronger than the one of the ratio of \PQb jets to inclusive jets. 
This is because gluons, which dominate the inclusive jet sample, have on average broader patterns than light-quark jets. 
The momentum balance \zg, as shown in Fig.~\ref{fig:b_vs_light} (right), is more asymmetric in \PQb jets with respect to the light-quark jets according to the simulation.

\section{Summary}\label{sec:summary}
This paper presents measurements of the substructure of bottom quark jets (\PQb jets) and of inclusive jets in proton-proton collisions at ${\sqrt{s} = 5.02\TeV}$ using data corresponding to an integrated luminosity of 301\pbinv collected in 2017 with the CMS experiment. 
The average number of simultaneous collisions per bunch crossing over the entire data-taking period was approximately 2.
The jets considered were initially clustered with the anti-\kt algorithm using a distance parameter of $R = 0.4$, and had transverse momentum ${100<\pt<120\GeV}$ and pseudorapidity $\abs{\eta}<2$. 
The substructure observables are calculated using the charged-particle constituents of the jets.
The groomed jet radius \rg, groomed momentum balance \zg, and the jet fragmentation function \zch are measured and corrected to the charged-particle level. 
Corrections include background subtraction, migration effects, and efficiency corrections.

A challenge in the interpretation of previous \PQb jet substructure measurements is that the \PQb hadron decay daughters are clustered in the jet on an equal footing with the hadrons that are produced from the transition from partons to hadrons. 
To better understand the gluon radiation pattern of \PQb jets, the \PQb hadron decay daughters need to be handled in a careful way experimentally. 
This measurement uses a novel algorithm to identify and cluster the \PQb hadron charged decay daughters of \PQb-tagged jets in a generic way, which allows for a cleaner interpretation of the measured substructure variable distributions in terms of the parton showering description. 

The contributions from gluon, light-quark, charm quark jets, and jets originating from gluon splitting ($\Pg \to \PQb \PAQb$) are subtracted from the measurements. 
The distributions are unfolded to the charged-particle level using unregularized unfolding. 
The correction to the charged-particle level accounts also for the removal of the particle-level biases that arise from the usage of \PQb tagging.

The charged-particle-level distributions have uncertainties of the order of 5\% and reaching up to 20\% in certain bins.
The dominant systematic uncertainty is related to the physics model (\PYTHIA{8} CP5 or \HERWIG{7} CH3) used in extraction of the corrections.
The groomed momentum balance distributions of inclusive jets and \PQb jets are well described by both \PYTHIA{8} CP5 and \HERWIG{7} CH3.
The description of the groomed jet radius, however, depends on the jet flavor and the physics model. 
The \HERWIG{7} CH3 prediction agrees well with the data in the case of inclusive jets, while a bigger discrepancy is observed for \PQb jets. 
The \PYTHIA{8} CP5 generator fails to describe either distribution.
The jet fragmentation function, on the other hand, is described better by \PYTHIA{8} CP5 at high values of \zch, \ie, for cases where the \PQb hadron carries a substantial amount of the jet \pt. 
The $\zch \approx 1$ region corresponds to \PQb jets without splittings that satisfy the soft-drop condition. 

Therefore, \zg, \rg, and \zch can be used simultaneously to constrain nonperturbative and perturbative aspects of \PQb jet substructure. 
The \rg distribution for \PQb jets is found to differ significantly from that of inclusive jets.
The suppression at small \rg for \PQb jets is consistent with the expectations of the dead-cone effect. 
The \PQb jets also have a more asymmetric \zg distribution than the inclusive jet reference, indicating that the \PQb quark carries a greater amount of momentum when emitting gluons.
The identified \PQb quark dead cone may serve as a control region for isolated medium-induced radiation in Pb-Pb collisions in a future measurement. 

\begin{acknowledgments}
We congratulate our colleagues in the CERN accelerator departments for the excellent performance of the LHC and thank the technical and administrative staffs at CERN and at other CMS institutes for their contributions to the success of the CMS effort. In addition, we gratefully acknowledge the computing centers and personnel of the Worldwide LHC Computing Grid and other centers for delivering so effectively the computing infrastructure essential to our analyses. Finally, we acknowledge the enduring support for the construction and operation of the LHC, the CMS detector, and the supporting computing infrastructure provided by the following funding agencies: SC (Armenia), BMBWF and FWF (Austria); FNRS and FWO (Belgium); CNPq, CAPES, FAPERJ, FAPERGS, and FAPESP (Brazil); MES and BNSF (Bulgaria); CERN; CAS, MoST, and NSFC (China); MINCIENCIAS (Colombia); MSES and CSF (Croatia); RIF (Cyprus); SENESCYT (Ecuador); ERC PRG, TARISTU24-TK10 and MoER TK202 (Estonia); Academy of Finland, MEC, and HIP (Finland); CEA and CNRS/IN2P3 (France); SRNSF (Georgia); BMFTR, DFG, and HGF (Germany); GSRI (Greece); NKFIH (Hungary); DAE and DST (India); IPM (Iran); SFI (Ireland); INFN (Italy); MSIT and NRF (Republic of Korea); MES (Latvia); LMTLT (Lithuania); MOE and UM (Malaysia); BUAP, CINVESTAV, CONACYT, LNS, SEP, and UASLP-FAI (Mexico); MOS (Montenegro); MBIE (New Zealand); PAEC (Pakistan); MES, NSC, and NAWA (Poland); FCT (Portugal); MESTD (Serbia); MICIU/AEI and PCTI (Spain); MOSTR (Sri Lanka); Swiss Funding Agencies (Switzerland); MST (Taipei); MHESI (Thailand); TUBITAK and TENMAK (T\"{u}rkiye); NASU (Ukraine); STFC (United Kingdom); DOE and NSF (USA).

\hyphenation{Rachada-pisek} Individuals have received support from the Marie-Curie program and the European Research Council and Horizon 2020 Grant, contract Nos.\ 675440, 724704, 752730, 758316, 765710, 824093, 101115353, 101002207, 101001205, and COST Action CA16108 (European Union); the Leventis Foundation; the Alfred P.\ Sloan Foundation; the Alexander von Humboldt Foundation; the Science Committee, project no. 22rl-037 (Armenia); the Fonds pour la Formation \`a la Recherche dans l'Industrie et dans l'Agriculture (FRIA) and Fonds voor Wetenschappelijk Onderzoek contract No. 1228724N (Belgium); the Beijing Municipal Science \& Technology Commission, No. Z191100007219010, the Fundamental Research Funds for the Central Universities, the Ministry of Science and Technology of China under Grant No. 2023YFA1605804, the Natural Science Foundation of China under Grant No. 12061141002, 12535004, and USTC Research Funds of the Double First-Class Initiative No.\ YD2030002017 (China); the Ministry of Education, Youth and Sports (MEYS) of the Czech Republic; the Shota Rustaveli National Science Foundation, grant FR-22-985 (Georgia); the Deutsche Forschungsgemeinschaft (DFG), among others, under Germany's Excellence Strategy -- EXC 2121 ``Quantum Universe" -- 390833306, and under project number 400140256 - GRK2497; the Hellenic Foundation for Research and Innovation (HFRI), Project Number 2288 (Greece); the Hungarian Academy of Sciences, the New National Excellence Program - \'UNKP, the NKFIH research grants K 131991, K 133046, K 138136, K 143460, K 143477, K 146913, K 146914, K 147048, 2020-2.2.1-ED-2021-00181, TKP2021-NKTA-64, and 2021-4.1.2-NEMZ\_KI-2024-00036 (Hungary); the Council of Science and Industrial Research, India; ICSC -- National Research Center for High Performance Computing, Big Data and Quantum Computing, FAIR -- Future Artificial Intelligence Research, and CUP I53D23001070006 (Mission 4 Component 1), funded by the NextGenerationEU program (Italy); the Latvian Council of Science; the Ministry of Education and Science, project no. 2022/WK/14, and the National Science Center, contracts Opus 2021/41/B/ST2/01369, 2021/43/B/ST2/01552, 2023/49/B/ST2/03273, and the NAWA contract BPN/PPO/2021/1/00011 (Poland); the Funda\c{c}\~ao para a Ci\^encia e a Tecnologia, grant CEECIND/01334/2018 (Portugal); the National Priorities Research Program by Qatar National Research Fund; MICIU/AEI/10.13039/501100011033, ERDF/EU, "European Union NextGenerationEU/PRTR", and Programa Severo Ochoa del Principado de Asturias (Spain); the Chulalongkorn Academic into Its 2nd Century Project Advancement Project, the National Science, Research and Innovation Fund program IND\_FF\_68\_369\_2300\_097, and the Program Management Unit for Human Resources \& Institutional Development, Research and Innovation, grant B39G680009 (Thailand); the Kavli Foundation; the Nvidia Corporation; the SuperMicro Corporation; the Welch Foundation, contract C-1845; and the Weston Havens Foundation (USA).
\end{acknowledgments}

\bibliography{auto_generated}

\appendix
\numberwithin{figure}{section}
\numberwithin{table}{section}
\section[The b hadron decay product identification and clustering]{The \PQb hadron decay product identification and clustering}\label{app:bid}
\begin{figure}[!ht]
    \centering
    \includegraphics[width=0.6\linewidth]{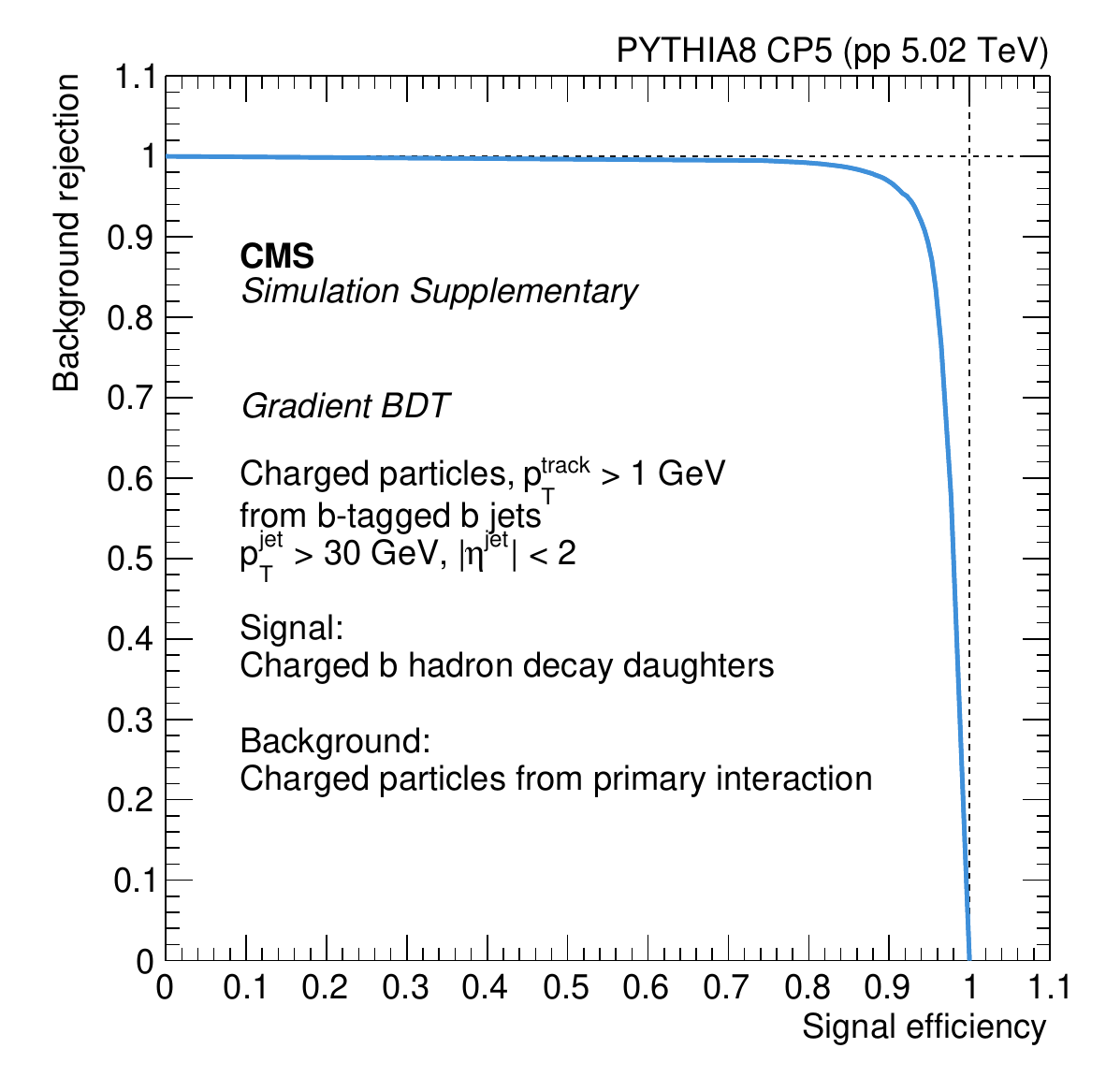}
    \caption{The background rejection rate versus the signal efficiency for the \PQb hadron decay product identification. The model is a gradient boosted decision tree with 11 input variables related to tracking and secondary vertex information. ``Background'' refers to charged particles coming from the primary interaction, while ``signal'' signifies charged particles resulting from a \PQb hadron decay. Dashed lines are drawn at 1 for each axis. The intersection of those lines is the optimal performance of a classifier.}
    \label{fig:roc_curve}
\end{figure}

\begin{figure}[!ht]
    \centering
    \includegraphics[width=0.49\linewidth]{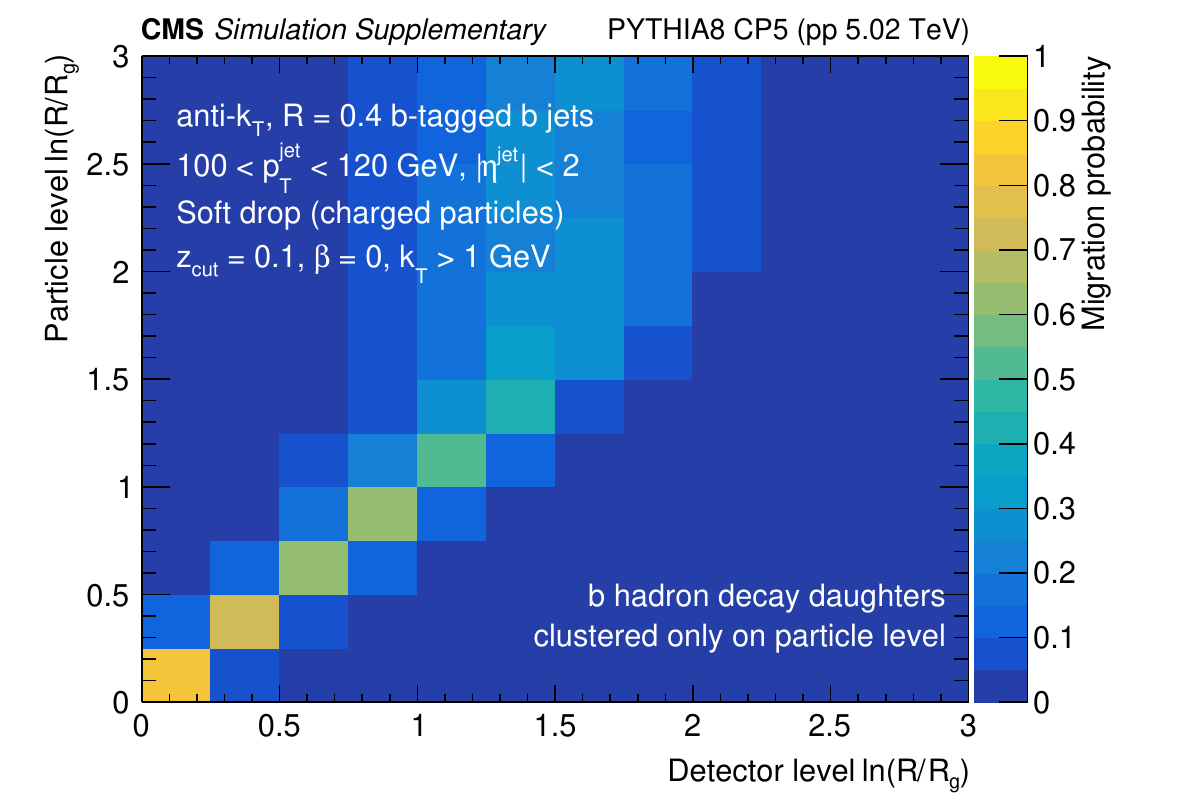}
    \includegraphics[width=0.49\linewidth]{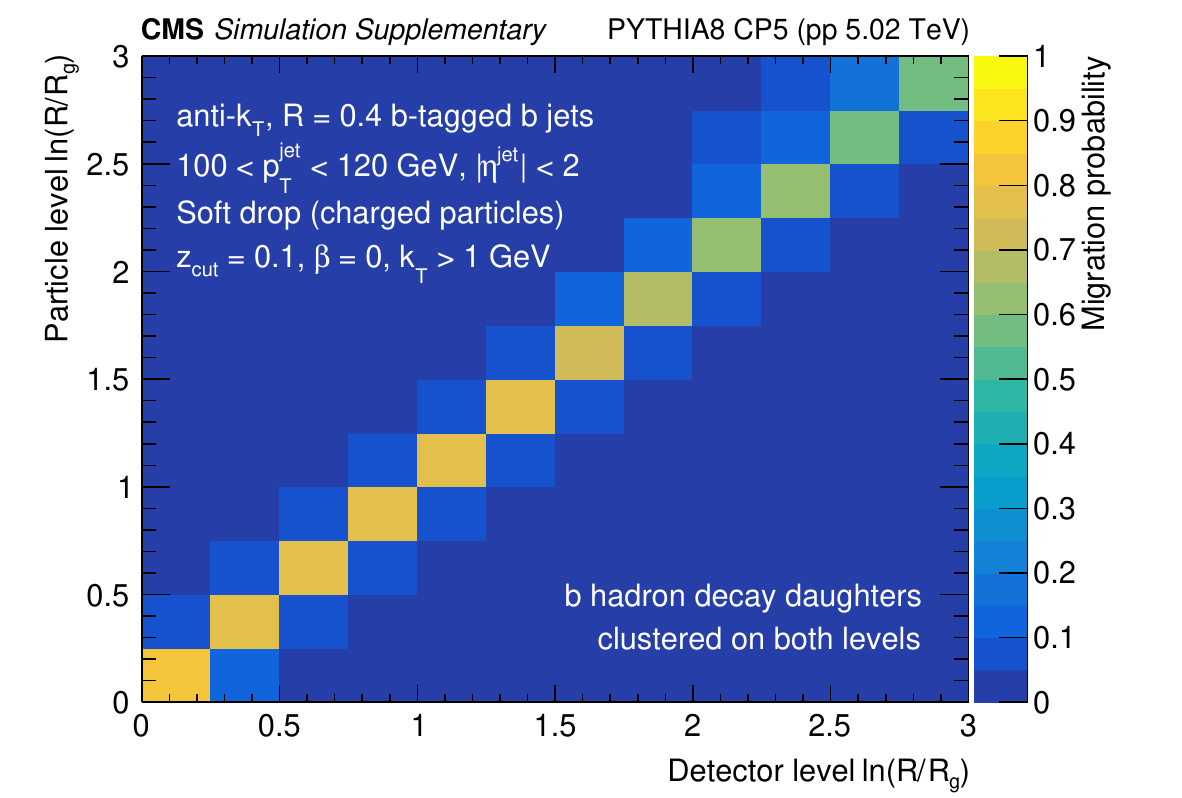}
    \caption{The migration matrices of \rg for \PQb jets. The particle level corresponds to clustered \PQb hadron decay products. In the left panel, the decay products are present during the iterative declustering at the detector level, while in the right panel, the decay products have been identified using the gradient boosted decision tree and they have been clustered together into the partially reconstructed \PQb hadron and replaced in the jet constituents.}
    \label{fig:aggr_vs_not_response}
\end{figure}
\cleardoublepage \section{The CMS Collaboration \label{app:collab}}\begin{sloppypar}\hyphenpenalty=5000\widowpenalty=500\clubpenalty=5000
\cmsinstitute{Yerevan Physics Institute, Yerevan, Armenia}
{\tolerance=6000
V.~Chekhovsky, A.~Hayrapetyan, V.~Makarenko\cmsorcid{0000-0002-8406-8605}, A.~Tumasyan\cmsAuthorMark{1}\cmsorcid{0009-0000-0684-6742}
\par}
\cmsinstitute{Institut f\"{u}r Hochenergiephysik, Vienna, Austria}
{\tolerance=6000
W.~Adam\cmsorcid{0000-0001-9099-4341}, J.W.~Andrejkovic, L.~Benato\cmsorcid{0000-0001-5135-7489}, T.~Bergauer\cmsorcid{0000-0002-5786-0293}, S.~Chatterjee\cmsorcid{0000-0003-2660-0349}, K.~Damanakis\cmsorcid{0000-0001-5389-2872}, M.~Dragicevic\cmsorcid{0000-0003-1967-6783}, P.S.~Hussain\cmsorcid{0000-0002-4825-5278}, M.~Jeitler\cmsAuthorMark{2}\cmsorcid{0000-0002-5141-9560}, N.~Krammer\cmsorcid{0000-0002-0548-0985}, A.~Li\cmsorcid{0000-0002-4547-116X}, D.~Liko\cmsorcid{0000-0002-3380-473X}, I.~Mikulec\cmsorcid{0000-0003-0385-2746}, J.~Schieck\cmsAuthorMark{2}\cmsorcid{0000-0002-1058-8093}, R.~Sch\"{o}fbeck\cmsAuthorMark{2}\cmsorcid{0000-0002-2332-8784}, D.~Schwarz\cmsorcid{0000-0002-3821-7331}, M.~Sonawane\cmsorcid{0000-0003-0510-7010}, W.~Waltenberger\cmsorcid{0000-0002-6215-7228}, C.-E.~Wulz\cmsAuthorMark{2}\cmsorcid{0000-0001-9226-5812}
% ...
\par}
\vskip\cmsinstskip
\dag:~Deceased\\
$^{1}$Also at Yerevan State University, Yerevan, Armenia\\
$^{2}$Also at TU Wien, Vienna, Austria\\
% ...
\end{sloppypar}
%%% END EDITABLE REGION %%%
% skeleton_end
\end{document}